\documentclass[preprint,onecolumn,nofootinbib]{revtex4}
\pdfoutput=1
\usepackage{float}
\usepackage{graphicx}
\usepackage{amsmath}
\usepackage{amsfonts}
\usepackage{amssymb,ulem}
\usepackage{color}%
\usepackage{dcolumn}
\usepackage{caption2}
\usepackage{subfigure}

\usepackage{MnSymbol,wasysym}
\usepackage{braket,diagbox}
\usepackage{eurosym}
\usepackage{calrsfs}
\usepackage[usenames,dvipsnames,svgnames]{xcolor}

\newcommand{\RNum}[1]{\uppercase\expandafter{\romannumeral #1\relax}}
\usepackage[colorlinks=true,linkcolor=blue,urlcolor=blue,filecolor=black,citecolor=red,
pdfstartview=FitV,pdftitle={},pdfsubject={},pdfkeywords={},pdfpagemode=None,bookmarksopen=true]{hyperref}

\begin{document}
\baselineskip=0.5 cm

\title{Photon regions, shadow observables and constraints from M87* of a charged rotating black hole}
\author{Yuan Meng}
\author{Xiao-Mei Kuang}
\email{xmeikuang@yzu.edu.cn}
\affiliation{Center for Gravitation and Cosmology, College of Physical Science and Technology, Yangzhou University, Yangzhou, 225009, China}
\author{Zi-Yu Tang}
\affiliation{School of Fundamental Physics and Mathematical Sciences, Hangzhou Institute for Advanced Study, UCAS, Hangzhou 310024, China}
\affiliation{School of Physical Sciences, University of Chinese Academy of Sciences, Beijing 100049, China}

\begin{abstract}
\baselineskip=0.45 cm
Inspired by the observations of supermassive black hole M87* in \emph{Event Horizon Telescope }(EHT) experiment, a remarkable surge in black hole physics is to use the black hole shadow's observables to distinguish general relativity (GR) and modified theories of gravity (MoG), which could also help to disclose the astrophysical nature of the center black hole in EHT observation. In this paper, we shall extensively carry out the study of a charged rotating black hole in conformal gravity, in which the term related with the charge has different falloffs from the usual Kerr-Newman (KN) black hole. We investigate the spacetime properties including the horizons, ergospheres and the photon regions; afterward, we show the boundary of black hole shadow and investigate its characterized observables. The features closely depend on the spin and charge parameters, which are compared with those in Kerr and KN black holes.  Then presupposing the M87* a charged rotating black hole in conformal gravity, we also constrain the black hole parameters via the observation constraints from EHT experiment.
We find that the constraints on the inferred circularity deviation, $\Delta C \lesssim 0.1$, and on the shadow axial ratio, $1< D_x \lesssim 4/3$, for the M87* black hole are satisfied for  the entire parameter space of the charged rotating black hole in conformal gravity. However, the shadow angular diameter $\theta_d = 42 \pm 3 \mu as$ will give upper bound on the parameter space. Our findings indicate that the current charged rotating black hole in conformal gravity could be a candidate for astrophysical black holes. Moreover, the EHT observation on the axial ratio $D_x$ may help us to distinguish Kerr black hole and the current charged rotating black hole in conformal gravity in some parameter space.

\end{abstract}


\maketitle
\tableofcontents
\section{Introduction}
Since Bardeen addressed that the shadow of the Kerr black hole would be distorted by the spin \cite{Bardeen1973} in contrast to a perfect circle for the Schwarzschild black hole \cite{Synge:1966okc}, the study on shadow of rotating black hole has been blooming with the motivation that
the trajectories of light near black hole and shadow are closely connected with the essential properties of the background theory of gravity. Thus, physicists could use shadow to unreveal the near horizon features of black hole  by analytical investigations or numerical simulation of their shadows \cite{Falcke:1999pj, Virbhadra:1999nm, Shen:2005cw,
Younsi:2016azx, Atamurotov:2013sca, Atamurotov:2015xfa, Amir:2017slq, Eiroa:2017uuq, Vagnozzi:2019apd,Long:2019nox,Long:2020wqj,Banerjee:2019nnj,Mishra:2019trb,Kumar:2020hgm,Qian:2021qow,Zeng:2020dco,Zeng:2021dlj,
Lin:2022ksb,Sun:2022wya,Cimdiker:2021cpz,Zhong:2021mty,Hou:2021okc,Cai:2021uov,Gan:2021pwu,Chang:2021ngy,
Wang:2021ara,Shaikh:2021cvl,Guo:2020blq} and therein.
Moreover, the size and distortion of shadow \cite{Hioki:2009na,Kumar:2018ple}, which could be calculated via the boundary of shadow,
has been widely investigated to estimate the black hole parameters in both GR and MoG, { with or without additional sources surrounding the black hole} \cite{Wei:2013kza,Allahyari:2019jqz,Tsupko:2017rdo,Cunha:2019dwb, Kumar:2020owy,Chen:2020aix,Brahma:2020eos,Belhaj:2020kwv,Badia:2021kpk,Lee:2021sws,Badia:2020pnh,Afrin:2021imp,Kumar:2019pjp,
Ghosh:2020spb,Bambi:2019tjh,Afrin:2021wlj,Jha:2021bue,Khodadi:2021gbc,Frion:2021jse,Roy:2021uye}. This direction could be seen as one aspect of black hole shadows to distinguish  GR and other theories of gravity, { or to acquire the information of the surrounding matter, though it was found that those theoretical features of shadow are usually not sufficient to distinguish black
holes in different theories or confirm the details of the surrounding matter}. More details about black hole shadows can be seen in the reviews \cite{Cunha:2018acu,Perlick:2021aok}.

More recently, the EHT collaboration captured the first image of the supermassive black hole M87* which makes the black hole shadow become a physical reality beyond theory \cite{EventHorizonTelescope:2019dse,EventHorizonTelescope:2019ths,EventHorizonTelescope:2019pgp}. The shadow of M87* from EHT observation has a derivation from circularity $\Delta C \lesssim 0.1$, a axis ratio $1< D_x \lesssim 4/3$ and the angular diameter $\theta_d=42\pm 3\mu as$. These observations are consistent with the image of Kerr black hole predicted from GR, but they cannot rule out Kerr or non-Kerr black holes in MoG. Thus, the EHT observations of shadow are then applied as an important tool to test black hole in strong gravitational field regime, as the observational data could be used to constrain the black hole parameters in MoG, and even to distinguish different theories of gravity
\cite{Cunha:2019ikd,EventHorizonTelescope:2021dqv,Khodadi:2020jij,Bambi:2019tjh,
Afrin:2021imp,Kumar:2019pjp,Ghosh:2020spb,Afrin:2021wlj,Jha:2021bue,Khodadi:2021gbc}.

In this work, we shall mainly study the aspects of shadows for a charged rotating black hole in conformal gravity characterized by the spin and charge parameters, in which the charge-related term has different falloffs from the usual KN black hole. We will show more details about this black hole geometry later in next section.
The charged rotating black hole here we consider was given in  \cite{Liu:2012xn} as a solution in conformal gravity with the Lagrangian
\begin{equation}\label{lagrangian}
L=\frac{1}{2}\gamma C^{\mu\nu\rho\sigma}C_{\mu\nu\rho\sigma}+\frac{1}{3}\gamma F^2
\end{equation}
which includes the Weyl-squared term minimally coupling to the Maxwell field. Here $C_{\mu\nu\rho\sigma}$ is the Weyl tensor and $F=dA$ is the strength of the Maxwell field.
Conformal gravity was pioneerly introduced by Weyl as an extension of GR \cite{Weyl:1918pdp} and later extensively considered by 't Hooft etc. in \cite{Mannheim:1988dj,Varieschi:2009vlp,tHooft:2010xlr,tHooft:2014swy} and therein.
The analysis of ghost instability and unitary of conformal gravity has been studied in \cite{Bender:2007wu,Mannheim:2000ka}.
Different from GR, in conformal gravity the dark matter or dark energy is not necessary to solve several cosmological and astrophysical problems,  and readers  can refer to \cite{Mannheim:2011ds} for more details on this symposium. In addition, Maldacena addressed that conformal gravity would reduce to Einstein gravity for a certain boundary condition and there could be a holographic connection between the two theories of gravity \cite{Maldacena:2011mk}.
Such advantageous features indicate that the contents in conformal gravity deserve to explore further. One natural direction is the black hole shadow,  as the recent progress on EHT experiment opens a new window to test the strong field regime.

The shadow boundary of Kerr-like metric in conformal gravity has been investigated in \cite{Mureika:2016efo}. Here, we
consider the charged rotating black hole geometry and extensively study the aspects of its shadow.  Starting from the null geodesics, we study the photon regions and then figure out the shadow boundary of the black hole. We also analyze the characterized observables, i.e. the shape, size and distortion of the shadows and argue the estimation of the black hole parameters from  given observables. Then we consider the M87* as the charged rotating black hole in conformal gravity and constrain the black hole parameters with the EHT observations.

The remaining parts of this paper are organized as follows.
In section \ref{sec:BG features}, we study the horizons, static limit and other spacetime properties of the charged rotating black hole in conformal gravity. We obtain the photon region by analyzing the null geodesics in section \ref{sec:photon region}, and in section \ref{sec:shadows} with the use of Cartesian coordinates, we show the shadow boundary with various values of the parameters for observers at finite distance. In section \ref{sec:parameter estimation}, we investigate the size and deformation of the black hole shadow for infinite distant  observer and address the parameter estimation by the shadow observables, from which we also calculate  the energy emission rate. In section \ref{sec:constraints}, by presupposing the M87* the current charged rotating black hole in conformal gravity, we constrain the black hole parameters from the EHT observations. The last section contributes to our closing remarks.

\section{The charged rotating black hole in conformal gravity}\label{sec:BG features}
Starting from \eqref{lagrangian}, a rotating charged  black hole in conformal gravity was constructed in \cite{Liu:2012xn} with the metric
\begin{eqnarray}\label{eq-metric}
ds^2=&&\Sigma\left(\frac{1}{\Delta_r}dr^2+d\vartheta^2\right)+\frac{1}{\Sigma}\left((\Sigma+a \chi)^2  \sin^2\vartheta-\Delta_r \chi^2\right)d\varphi^2 \nonumber\\
&&+\frac{2}{\Sigma}\left(\Delta_r\chi-a(\Sigma+a\chi) \sin^2\vartheta\right)dt d\varphi-\frac{1}{\Sigma}\left(\Delta_r-a^2 \sin^2\vartheta\right)dt^2~,
\end{eqnarray}
where
\begin{eqnarray}
\Sigma=r^2+a^2 \cos^2\vartheta, ~~\chi=a \sin^2\vartheta,~~\Delta_r=r^2-2mr+a^2+\frac{\beta r^3}{6m}~.
\end{eqnarray}
Here $m$, $\beta=p^2+q^2$ and $a$ are the mass, charge and rotating parameters, respectively. When the charge parameter vanishes, the metric reduces to the well-known Kerr black hole. This black hole is different from the usual KN black hole where the charge term in $\Delta_r$ is simply a constant $\beta$, instead of the cube term $\beta r^3/6m$ in  current conformal gravity. It is noted that comparing the expression of the rotating charged solution in \cite{Liu:2012xn}, here we focus on the case with the integral constant $\Lambda$ being zero \footnote{We appreciate professor Hai-Shan Liu reminding us this point. }.

\subsection{Black hole horizons}

\begin{figure} [H]
\centering
\includegraphics[width=2in]{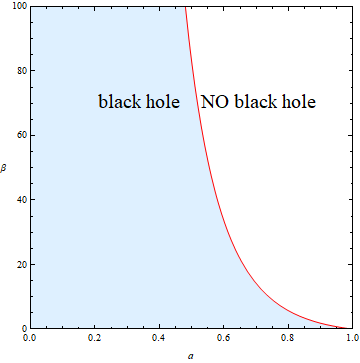}\hspace{2cm}
\includegraphics[width=2in]{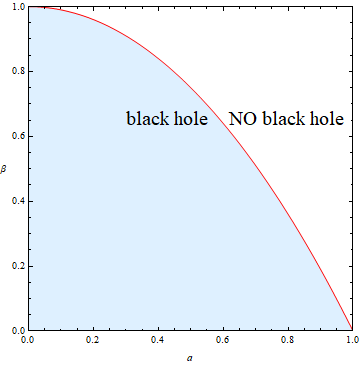}
   \caption{The parameter space $(a,\beta)$ of the charged rotating black hole in conformal gravity (left) and KN black hole (right). The red curves correspond to the extremal case that separating black holes (blue regions) from naked singularities (white regions). }   \label{fig:a-beta}
\end{figure}

It is known that $\Sigma\neq0$ and $g^{rr}=0$ could determine the black hole horizons, which correspond to the positive roots of
\begin{equation}\label{eq-grr0}
\Delta_r=r^2-2mr+a^2+\frac{\beta r^3}{6m}=0.
\end{equation}
There are three roots to the above equation. Depending on $m,a$ and $\beta$, the three roots could have two real positive values, one real positive value or no real positive value. The three cases correspond to that the metric \eqref{eq-metric} describes a non-extremal black hole with event horizon $(r_+)$ and Cauchy horizon $(r_-)$, extremal black hole with event horizon $r_{ex}=r_+=r_-$ and no black hole sector, respectively.
When $\beta$ is smaller than the critical value from the extremal condition
\begin{equation}\label{eq-extremalCondition}
\beta_{ex}=\frac{4 \left(8 m^4-9 a^2 m^2+\sqrt{m^2 \left(4 m^2-3 a^2\right)^3}\right)}{9 a^4}~,
\end{equation}
the metric describes a non-extremal black hole with  $0<r_-<r_+$. A naked singularity emerges when $\beta>\beta_{ex}$ because in this case none of the three roots is real positive.
Besides, as $\beta=0$, the horizons $r_{\pm}$ reduce to be $m\pm \sqrt{m^2-a^2}$ with $|a|\le m$ (Kerr case). The extremal value $\beta_{ex}$ is different from that for KN black hole ($\beta_{ex}^{KN}=m^2-a^2$). While for $a\to 0$ we have $\beta_{ex}\to +\infty$, which indicates the black hole is always non-extremal, in contrast to a finite value $\beta_{ex}^{KN}=m^2$ for Reissner-Nordstrom (RN) black hole.
The above scenarios in $(a,\beta)$ parameter space is shown in FIG.  \ref{fig:a-beta}, where the case for KN black hole is also present for comparison.

Note that here all parameters could be re-scaled to be dimensionless, depending on their dimensions related with $m$, for example, $a/m, r/m, \beta $ are dimensionless quantities. All the numerical exhibition of the quantities in this work  denote the dimensionless ones, and for simplicity, we will set $m=1$ in the calculations unless we reassign.

\begin{figure} [H]
\centering
\includegraphics[width=2.5in]{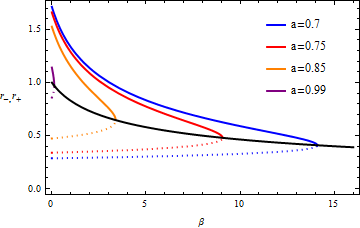}\hspace{1cm}
\includegraphics[width=2.5in]{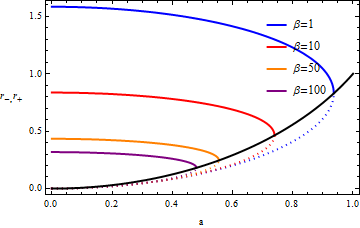}
   \caption{The event horizon $r_+$ (solid curve) and Cauchy horizon $r_-$ (dashed curve) are plotted with various values of $a$ and $\beta$. While the solid black curve represents the extremal case where the event horizon and Cauchy horizon coincide with each other.}   \label{fig:horizon}
\end{figure}

The explicit dependencies of the horizons on the parameters are shown in FIG.  \ref{fig:horizon}. It is obvious that as $\beta$ or $a$ increases, $r_{+}$ decreases while $r_{-}$ increases; as the extremal condition \eqref{eq-extremalCondition} is satisfied, $r_{+}$ and $r_{-}$ converge to $r_{ex}$ which decreases as $\beta$ increases but increases as $a$ increases (see the solid black curves). Here the effects of the charge and spin parameters on $r_{+}$ and $r_{-}$ are similar with that in KN spacetime, where, however the extremal horizon is $r_{ex}^{KN}=m$ independent of the charge and spin parameters.

\subsection{Static limit surface}
\begin{figure} [h]
\centering
\includegraphics[width=2in]{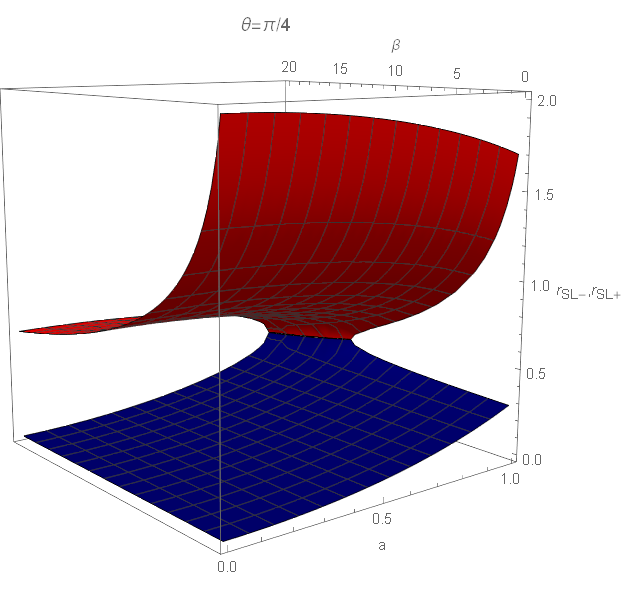}
\includegraphics[width=2in]{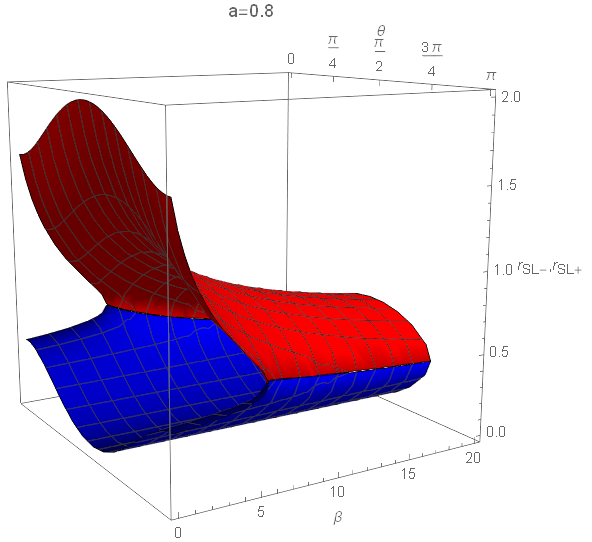}
\includegraphics[width=2in]{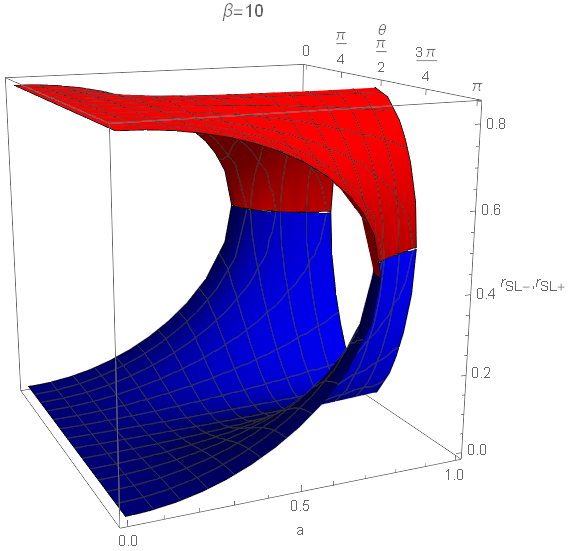}
   \caption{The static limit surfaces are plotted with fixed $\vartheta$, $a$ and $\beta$ respectively, where the red surface denotes  $r_{SL_{+}}$ while the blue surface represents for  $r_{SL_{-}}$.}   \label{fig:rSL}
\end{figure}

For a rotating black hole, the event horizon of the black hole does not coincide with the static limit surface, at which the asymptotical time translational Killing vector is null and therefore we have
\begin{equation}
g_{tt}=-\frac{1}{\Sigma}\left(\Delta_r-a^2 \sin^2\vartheta\right)=0.
\end{equation}
Depending on the values of $a,\beta$ and $\vartheta$, the roots to the above equation have three cases: no real positive root, a double real positive root and two real positive roots. We denote the real positive roots as
$r_{SL_{-}}$ and $r_{SL_{+}}$ with $r_{SL_{-}}<r_{SL_{+}}$. { The explicit expressions of the solutions are so complicated that we do not show them here, instead, we plot their behaviors in FIG. \ref{fig:rSL}. From the figures we can see that, there exist at least one border on which the two static limit surfaces coincide $r_{SL_{-}}=r_{SL_{+}}$, i.e. the extremal case with one real positive root.}

Here we will not explicitly describe the dependence of $r_{SL_{\pm}}$ on the parameters $a,\beta$ and $\vartheta$.
{What we really want to show is that the ergoregion of this rotating black hole is bounded between $r_+<r<r_{SL+}$ and $r_-<r<r_{SL-}$, in which the timelike killing vector becomes spacelike ($g_{tt}>0$). Particularly, when $\Sigma=0$, which requires both $r=0$ and $a \cos\vartheta=0$, the spacetime has a true physical singularity. Apart from this ring singularity, the sphere $r=0$ is regular. Besides, for $g_{\varphi\varphi}<0$, the spacetime violates the causality condition, because of the closed timelike curves. More detailed exhibitions of the horizons, ergoregions, singularity and causality violating regions will be present later together with the photon regions.}

\section{Null geodesics and photon regions}\label{sec:photon region}
{The light propagation near a black hole has important significance in both theoretical physics and astrophysics, particularly the circular orbits. For photons, the circular orbits outside the event horizon of a black hole are usually unstable. This indicates that a slight perturbation can make the photons fall into the black hole, or escape to infinity, the latter can constitute a photon ring that confines the black hole image for observers at a distant. Therefore, we start from the geodesics of the photons, to analyze the photon regions and the shadow images in the charged rotating black hole spacetime \eqref{eq-metric} in conformal gravity.

We first consider the particles with mass $\mu$, the Lagrangian of which writes $\mathcal{L}=\frac{1}{2}g_{\mu\nu}\dot{x}^{\mu}\dot{x}^\nu$. Here the dot represents the derivative with respect to the affine parameter $\lambda$ which relates to the proper time via $\tau=\lambda \mu$. Following \cite{Carter:1968rr}, we introduce the Hamilton-Jacobi equation
\begin{equation}
\mathcal{H}=-\frac{\partial S}{\partial \lambda}=\frac{1}{2}g_{\mu\nu}\frac{\partial S}{\partial x^{\mu}}\frac{\partial S}{\partial x^{\nu}}=-\frac{1}{2}\mu^2,
\label{Lagrangian}
\end{equation}
where $\mathcal{H}$ and $S$ are the canonical Hamiltonian and the Jacobi action. With the conserved quantities
\begin{eqnarray}
E:=-\frac{\partial S}{\partial t}=-g_{\varphi t}\dot{\varphi}-g_{tt}\dot{t},~~~\mathrm{and}~~~
L_z:=\frac{\partial S}{\partial \varphi}=g_{\varphi\varphi}\dot{\varphi}+g_{\varphi t}\dot{t}~,
\label{momentum}
\end{eqnarray}
the Jacobi action can be separated as
\begin{equation}
S=\frac{1}{2}\mu^2\lambda-Et+L_z\varphi+S_r(r)+S_{\vartheta}(\vartheta),
\label{action}
\end{equation}
where  $E$, $L_z$ are the constants of motion associated with the  energy and angular momentum of the particle, respectively.

Then focusing on the photons ($\mu=0$), we obtain four first-order differential equations for the geodesic motions}
\begin{eqnarray}
\dot{t}&=&\frac{\chi(L_z -E \chi)}{\Sigma  \sin^2\vartheta}+\frac{(\Sigma+a \chi)((\Sigma+a \chi)E-a L_z )}{\Sigma \Delta_r}, \label{eq-motion2}\\
\dot{\varphi}&=&\frac{(L_z -E \chi)}{\Sigma \sin^2 \vartheta}+\frac{ a(E(a \chi+\Sigma)-a L_z )}{\Sigma \Delta_r}, \label{eq-motion1}\\
\Sigma^2 \dot{\vartheta}^2&=& K-\frac{(E \chi-L_z )^2}{\sin^2\vartheta}=:\Theta(\vartheta),\label{eq-motion3} \\
\Sigma^2 \dot{r}^2&=&((\Sigma+a \chi)E-a L_z )^2-\Delta_r K=:R(r), \label{eq-motion4}
\end{eqnarray}
where $K$ is Carter constant. Comparing to the complete solution to the above equations, we are more interested in the photon region, which is filled by the null geodesics staying on a sphere. For convenience, we introduce the abbreviations
\begin{equation}
L_E \equiv \frac{L_z}{E},~~~~~K_E \equiv \frac{K}{E^2}.
\end{equation}

The spherical orbits require $\dot{r}=0$ and $\ddot{r}=0$, which can be fulfilled by $R(r)=0$ and $R'(r)=0$ according to \eqref{eq-motion4}. Subsequently, the constants of motion $K_E$ and $L_E$ are given as
\begin{equation}
K_E=\frac{16r^2 \Delta_r}{(\Delta'_r)^2},~~~~a L_E =(\Sigma+a \chi)-\frac{4r \Delta_r}{\Delta'_r}, \label{eq-ke}
\end{equation}
where the prime denotes the derivative to $r$.
Substituting the above expression into \eqref{eq-motion3}, we find that its non-negativity could give us the condition for the photon region
\begin{equation}\label{eq-photonregion}
(4r \Delta_r-\Sigma \Delta'_r)^2\leq16 a^2 r^2 \Delta_r \sin^2\vartheta.
\end{equation}
In this region, for each point with coordinates $(r_p,\vartheta_p)$, there is a null geodesic staying on the sphere $r=r_p$, along which $\vartheta$ can oscillate between the extremal values determined by the equality in \eqref{eq-photonregion}, while the $\varphi$ is governed by \eqref{eq-motion1}.
With respect to radial perturbations, the spherical null geodesic at $r=r_p$ could be either unstable or stable depending on the sign of $R''(r_p)$ which can be derived from \eqref{eq-motion4} and \eqref{eq-ke} as
\begin{equation}
\frac{R''(r)}{8E^2}(\Delta'_r)^2=2r\Delta_r \Delta'_r+r^2(\Delta'_r)^2-2r^2 \Delta_r \Delta''_r.
\end{equation}
The condition $R''(r_p)>0$ means it is unstable while $R''(r_p)<0$ indicates the stability.

\begin{figure}[H]
\centering
\subfigure[$\beta=0$]
{\begin{minipage}[b]{0.22\textwidth}
      \centering
      \includegraphics[width=0.75\textwidth]{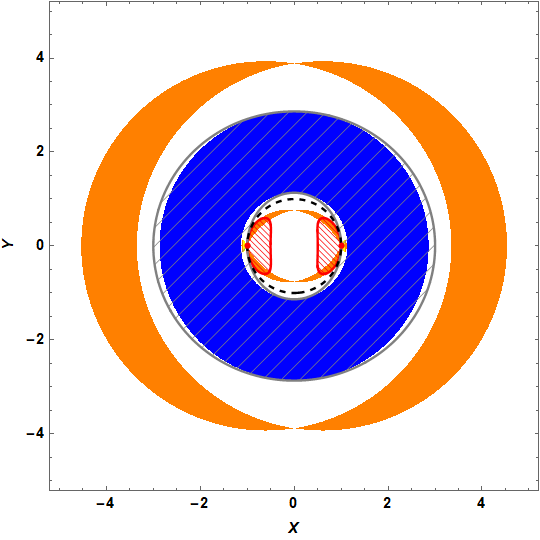}\\
      \includegraphics[width=0.75\textwidth]{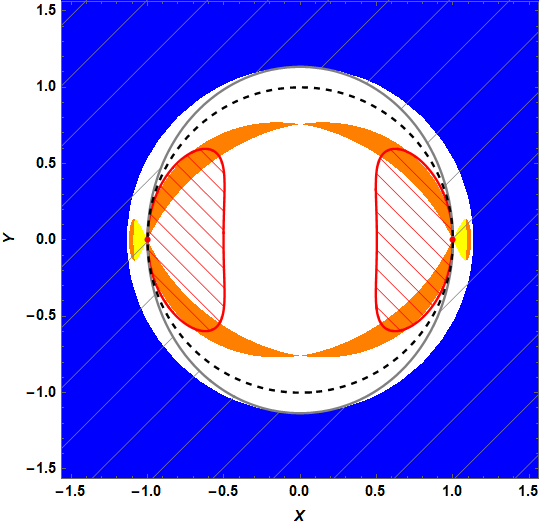}
 \end{minipage}
}
\subfigure[$\beta=0.3\beta_{ex}$]
{\begin{minipage}[b]{0.22\textwidth}
      \centering
      \includegraphics[width=0.75\textwidth]{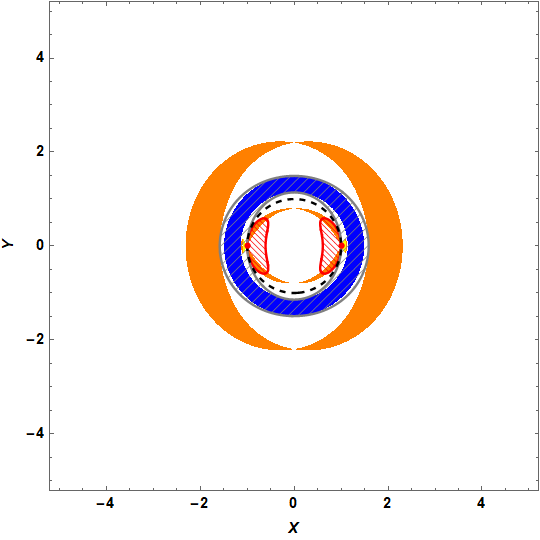}\\
      \includegraphics[width=0.75\textwidth]{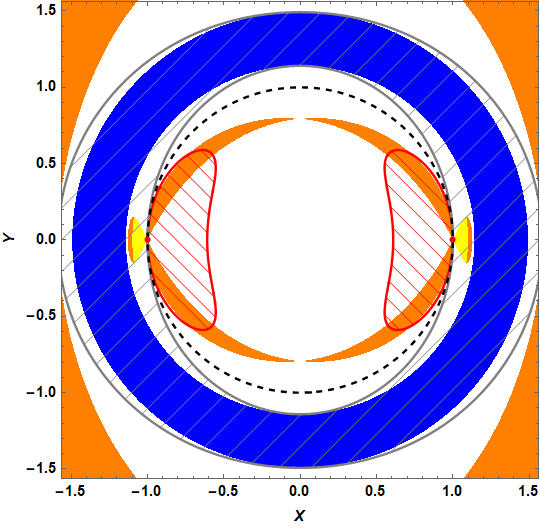}
 \end{minipage}
}
\subfigure[$\beta=0.6\beta_{ex}$]
{ \begin{minipage}[b]{0.22\textwidth}
      \centering
      \includegraphics[width=0.75\textwidth]{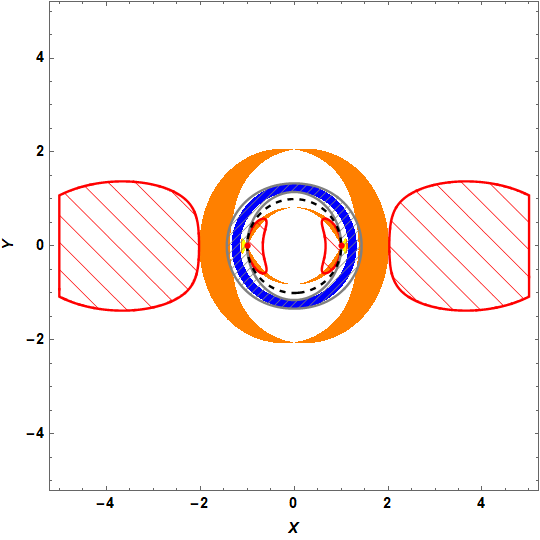}\\
      \includegraphics[width=0.75\textwidth]{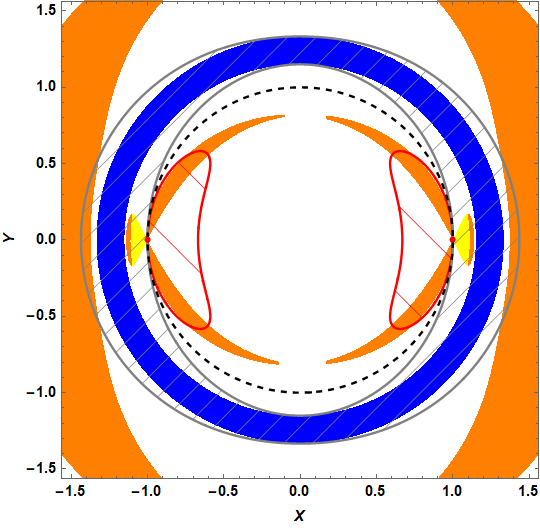}
   \end{minipage}
}
\subfigure[$\beta=\beta_{ex}$]
{  \begin{minipage}[b]{0.22\textwidth}
       \centering
       \includegraphics[width=0.75\textwidth]{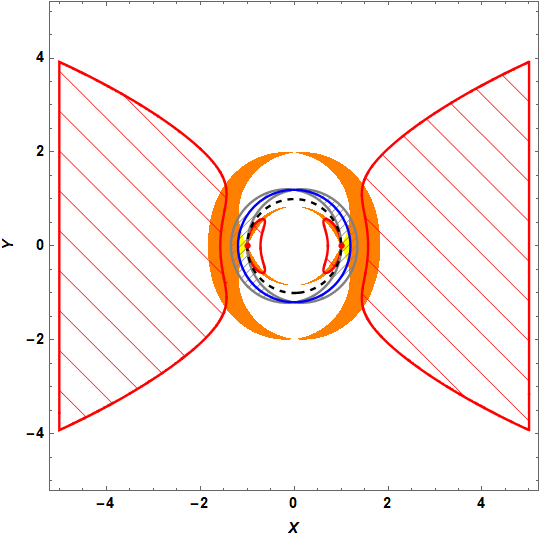}\\
       \includegraphics[width=0.75\textwidth]{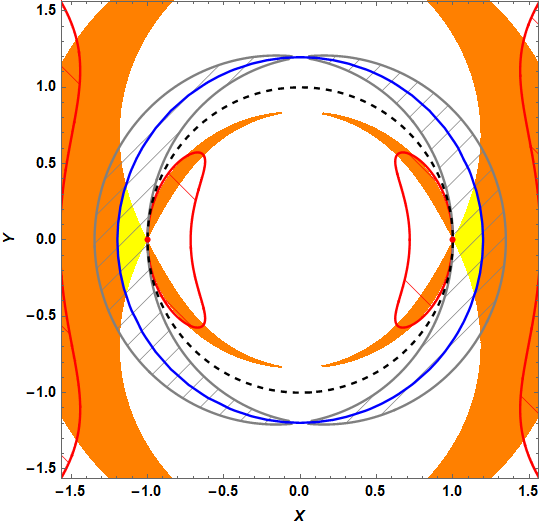}
    \end{minipage}
}
\caption{The photon regions with $a=0.5$, accompany with the unidirectional membrane region, the ergosphere region and the causality violation region. The plots in the bottom show a magnified inner part.}
\label{fig:photon region a=0.5}
\end{figure}

\begin{figure}[H]
\centering
\subfigure[$\beta=0$]
{\begin{minipage}[b]{0.22\textwidth}
      \centering
      \includegraphics[width=0.8\textwidth]{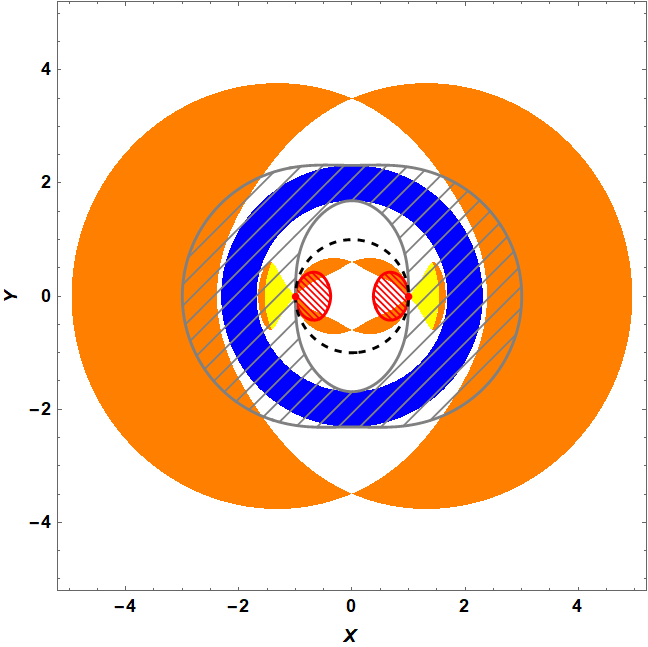}\\
 \end{minipage}
}
\subfigure[$\beta=0.3\beta_{ex}$]
{\begin{minipage}[b]{0.22\textwidth}
      \centering
      \includegraphics[width=0.8\textwidth]{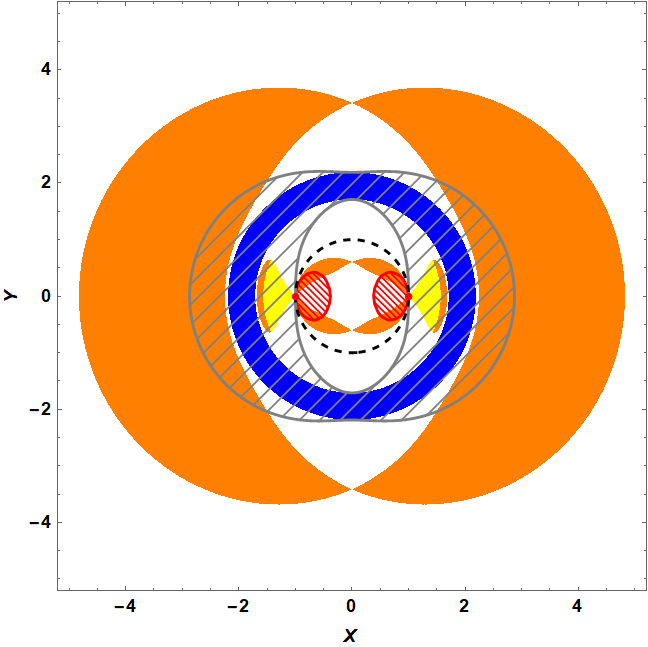}\\
 \end{minipage}
}
\subfigure[$\beta=0.6\beta_{ex}$]
{ \begin{minipage}[b]{0.22\textwidth}
      \centering
      \includegraphics[width=0.8\textwidth]{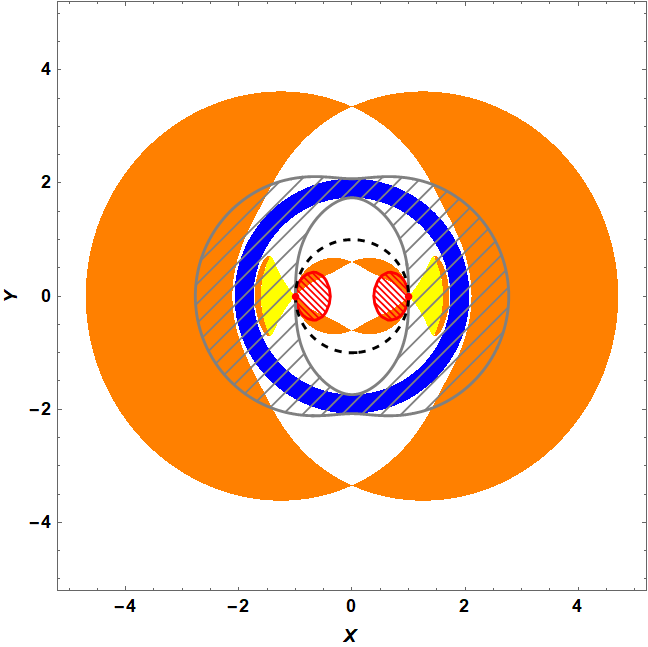}\\
   \end{minipage}
}
\subfigure[$\beta=\beta_{ex}$]
{  \begin{minipage}[b]{0.22\textwidth}
       \centering
       \includegraphics[width=0.8\textwidth]{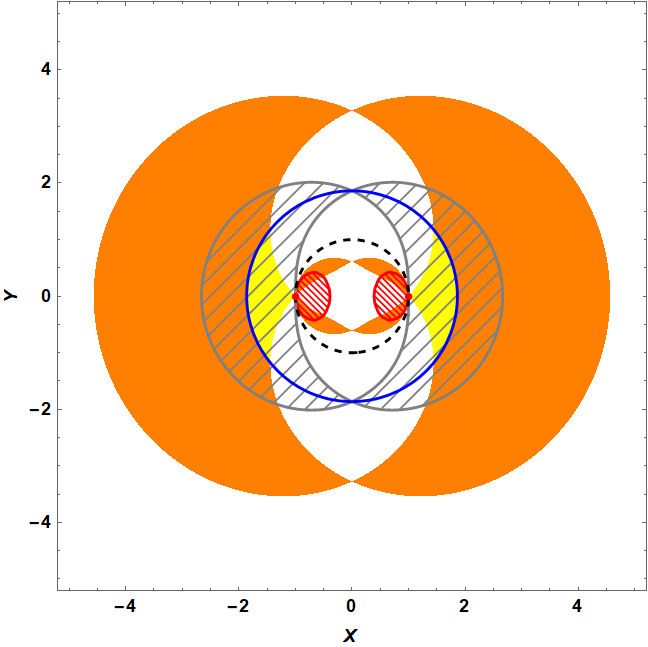}\\
    \end{minipage}
}
\caption{The photon regions with $a=0.95$, accompany with the unidirectional membrane region, the ergosphere region and the causality violation region.}
\label{fig:photon region a=0.95}
\end{figure}

The photon regions of the charged rotating black hole in conformal gravity are shown in $(r,\vartheta)$ plane, see FIG. \ref{fig:photon region a=0.5} and FIG. \ref{fig:photon region a=0.95}, where the unstable photon orbits (the \includegraphics[width=0.5cm,height=0.25cm]{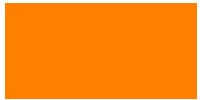} region) and stable photon orbits (the \includegraphics[width=0.5cm,height=0.25cm]{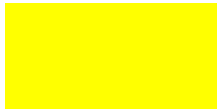} region) are distinguished. Here we plot the whole range of the spacetime in $r$ direction and use two different scales, following \cite{Grenzebach:2014fha}.  The radial coordinate has been scaled as $m\exp{(r/m)}$ in the region $r<0$, while scaled as $r+m$ in the region $r>0$, hence we use the black dashed circle to denote the throat at $r=0$.
Moreover, the \includegraphics[width=0.5cm,height=0.25cm]{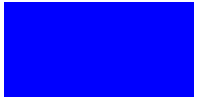} region represents $\Delta r \leq 0$ and its boundaries indicate the black hole horizons. The \includegraphics[width=0.5cm,height=0.25cm]{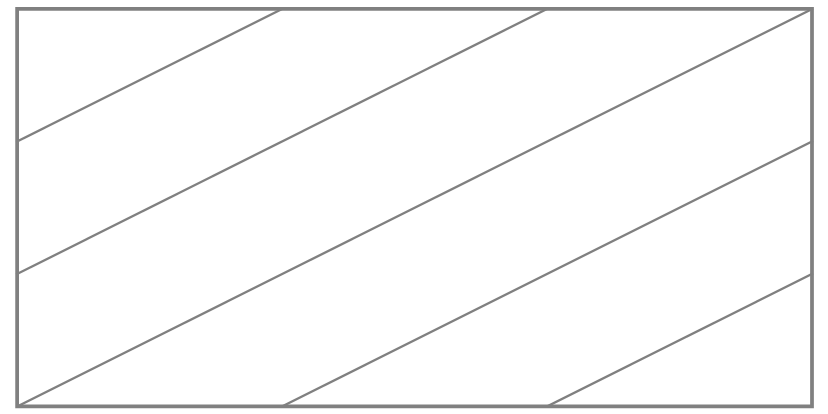} region and \includegraphics[width=0.5cm,height=0.25cm]{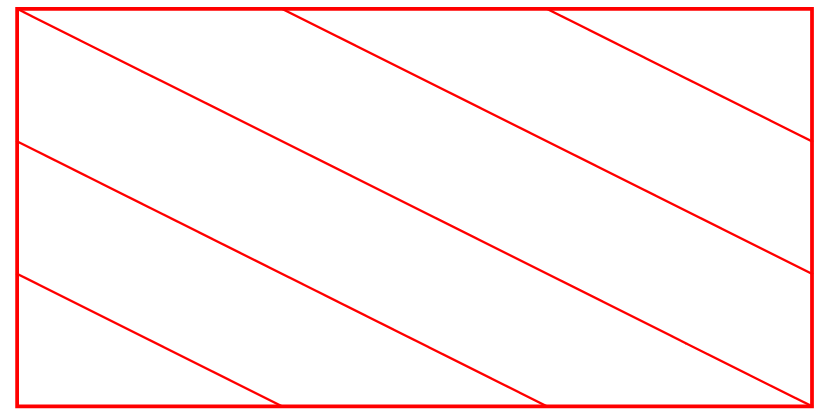} region represent the ergosphere and the causality violating regions, respectively. Besides, the \includegraphics[width=0.3cm,height=0.3cm]{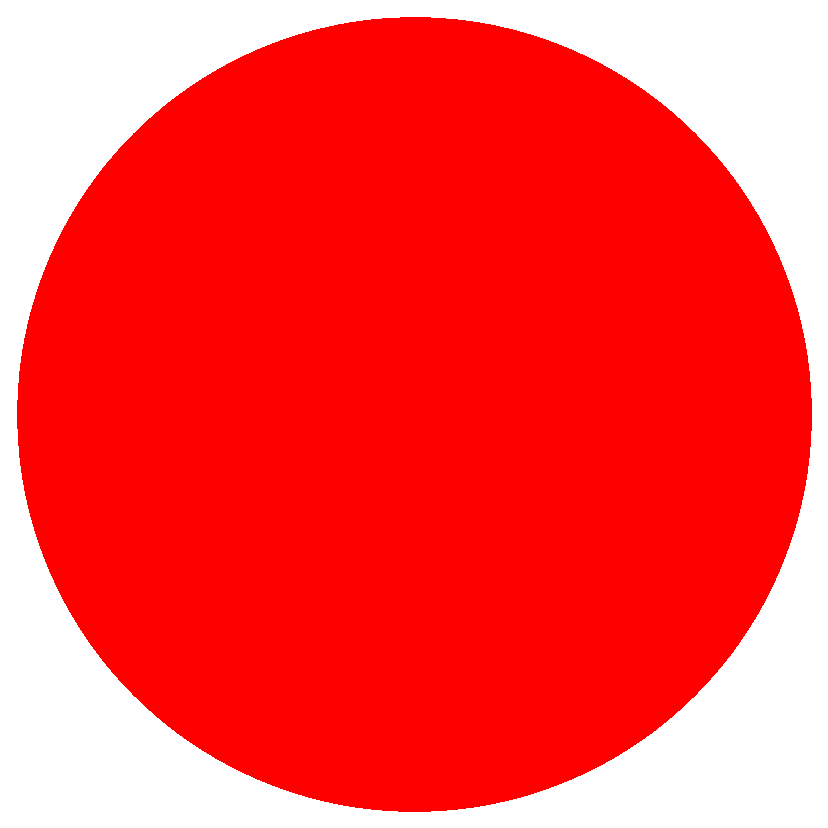} shows the singularity.

In the figures, we fix $a=0.5$ and $0.95$ respectively and change $\beta=\sharp\beta_{ex}$ where $\sharp\in{(0,0.3,0.6,1)}$ and $\beta_{ex}$ is the corresponding extremal value  \eqref{eq-extremalCondition}.
As in Kerr black hole \cite{Grenzebach:2014fha}, we see an exterior photon region outside the outer horizon and an interior photon region inside the inner horizon, which are symmetric with respect to the equatorial plane. All photon orbits are unstable in the exterior photon region while there exists both stable and unstable orbits in the interior photon region. The exterior and interior photon regions enlarge as $a$ increases but shrink as $\beta$ increases. Moreover, the dependence of the unidirectional membrane region and the ergosphere region on the black hole parameters are also obvious here and consistent with the analysis in the previous section. Also, the causality violation region lying to the side of negative $r$  always exists, and for small $a$ and large enough $\beta$, we see an additional causality violating region which is symmetric and extends from the the outer horizon to an finite region depending on $\beta$.



\section{Black hole shadows}\label{sec:shadows}
Since the photon region determines the boundary of the black hole shadow, we then go on to construct the shadow of the charged rotating black hole in conformal gravity.

\subsection{Coordinates setup}
\begin{figure} [H]
{\centering
\includegraphics[width=3.5in]{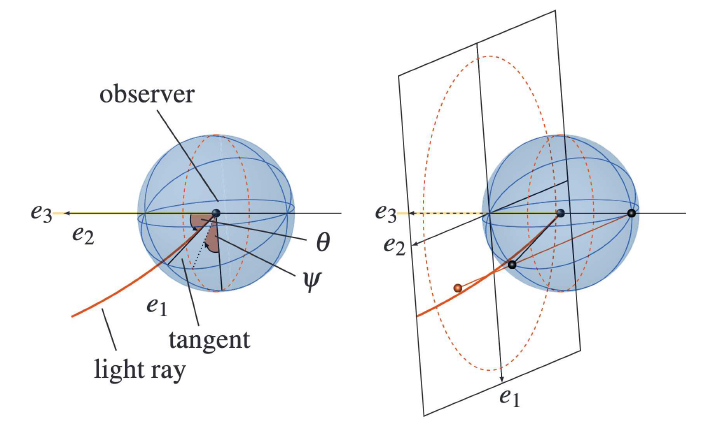}\hspace{0.5cm}
   \caption{This picture is taken from \cite{Perlick:2021aok}. The left picture shows the definition of the celestial coordinates $\theta$ and $\psi$ on the observer's sky. The right picture shows the stereographic projection of the celestial sphere onto a plane.}   \label{fig:cele-coordinate}}
\end{figure}

For light rays issuing from the position of an observer into the past, the initial direction is determined by two angles in the observer's sky, a colatitude angle and an azimuthal angle. Then we consider an observer at position $(r_o,\vartheta_o)$ in the Boyer-Lindquist coordinates. To fix the boundary of shadow, we choose an orthonormal tetrad \cite{Grenzebach:2014fha}
\begin{equation}\label{eq-tetrad}
\begin{split}
e_0=\frac{(\Sigma+a \chi)\partial t+ a \partial \varphi}{\sqrt{\Sigma \Delta r}}\Big\mid_{(r_o,\vartheta_o)},~~~~ e_1=\sqrt{\frac{1}{\Sigma}}\partial \vartheta\Big\mid_{(r_o,\vartheta_o)},\\ e_2=-\frac{ \partial \varphi+\chi \partial t}{\sqrt{\Sigma}\sin \vartheta}\Big\mid_{(r_o,\vartheta_o)},~~~~~ e_3=-\sqrt{\frac{\Delta _r}{\Sigma}}\partial r\Big\mid_{(r_o,\vartheta_o)},
\end{split}
\end{equation}
at the observation event in the domain of outer communication. In this set of tetrad, $e_0$ is treated as the four velocity of the observer and $e_3$ represents the spatial direction towards the center of the black hole, and $e_0\pm e_3$ are tangential to the principal null congruences of our background metric. In this way, a linear combination of $e_{i}$ is tangent to {a light ray $s (\lambda)=(t(\lambda),r(\lambda),\vartheta(\lambda),\varphi(\lambda))$, such that we have
\begin{equation}\label{eq-vector}
\partial_\lambda=\dot{r}\partial_r+\dot{\vartheta}\partial_{\vartheta}
+\dot{\varphi}\partial_{\varphi}+\dot{t}\partial_t=\alpha_s(-e_0+\sin\theta\cos\psi e_1+
\sin\theta\sin\psi e_2+\cos\theta e_3)
\end{equation}
}
where the scalar factor can be determined by inserting \eqref{eq-tetrad} into \eqref{eq-vector}  as
\begin{equation}
\alpha_s=\frac{aL_z-(\Sigma+a\chi)E}{\sqrt{\Sigma\Delta_r}}\Big\mid_{(r_o,\vartheta_o)},
\end{equation}
and it is easy to see that the direction $\theta=0$ points to the black hole.
Moreover, here we have introduced $\theta$ and $\psi$ which are the aforementioned two angles, i.e. the celestial coordinates in the observer's sky, see the left picture of FIG.  \ref{fig:cele-coordinate}. Further comparing the coefficients of $\partial_t$ and $\partial_r$, we find that
\begin{eqnarray}\label{eq-Cartesian1}
\sin\psi=\frac{ L_E-\chi}{\sqrt{K_E}\sin\vartheta}\Big\mid_{\vartheta=\vartheta_{o}}, ~~~\sin\theta=\frac{\sqrt{\Delta_r K_E}}{\Sigma+a \chi-a L_E }\Big\mid_{r=r_{o}}.
\end{eqnarray}
Since the boundary of shadow could correspond to the light rays which infinity approach a spherical null geodesic, so such light ray must have the same $K_E$ and $L_E$ as the limiting spherical null geodesic, so in \eqref{eq-Cartesian1}, we have
\begin{eqnarray}\label{eq-ke2}
K_E=\frac{16 r^2 \Delta_r}{(\Delta'_r)^2}\Big\mid_{r=r_p},~~~ a L_E =(\Sigma+a \chi)-\frac{4r \Delta_r}{\Delta'_r}\Big\mid_{r=r_p}.
\end{eqnarray}
where $r_p$ is the radius coordinate of the limiting spherical null geodesic.

Therefore, the boundary of the black hole shadow depends on $r_p$ in the form of $(\theta(r_p),\psi(r_p))$. Since the points $(\theta,\psi)$ and $(\theta,\pi-\psi)$ have the same $K_E$ and $L_E$, so the shadow is symmetric with respect to the horizontal axis. And for $a>0$, $\theta$ reaches its maximal and minimal value along the boundary curve at $\psi=-\pi/2$ and $\psi=\pi/2$, respectively, which could give us the corresponding $r_p^{max}$ and $r_p^{min}$. Putting \eqref{eq-ke2} into \eqref{eq-Cartesian1} with $\psi=\mp \pi/2$, $r_p^{max/min}$ can be solved via
\begin{equation}
(4r \Delta_r-\Sigma \Delta'_r)\mp 4 a r \sqrt{\Delta_r} \sin\vartheta \Big\mid_{(r=r_p,\vartheta=\vartheta_o)}=0.
\end{equation}
Note that for $a=0$, the above method that parameterizes the shadow boundary by $r_p$ does not work.

Then following \cite{Grenzebach:2014fha}, one could apply the stereographic projection (see the right picture of FIG.  \ref{fig:cele-coordinate}) to transform the celestial coordinates $(\theta(r_p),\psi(r_p))$ into the standard Cartesian coordinates  $(X(r_p),Y(r_p))$
\begin{equation}
\begin{split}
X{(r_p)}=-2\tan \left(\frac{\theta{(r_p)}}{2}\right)\sin\psi{(r_p)}, \\ Y{(r_p)}=-2\tan \left(\frac{\theta{(r_p)}}{2}\right)\cos\psi{(r_p)} .
\end{split}
\label{eq-Cartesian}
\end{equation}
Then we can figure out the boundary of the shadow on a two-dimensional plane, observed by our chosen observer with four-velocity $e_0$. Note that the range of the inclination angle is $\vartheta_o\in [0, \pi]$, and $\vartheta_o=0 (\pi)$ corresponds to the observer in north (south) direction while  $\vartheta_o= \pi/2$ corresponds to the observer at equatorial plane of the black hole. {Due to the symmetry, we shall consider $\vartheta_o\in [0, \pi/2]$ in the following study.}

\subsection{Shadow for observers at finite distance}
Firstly, we consider the observer located at finite distance with position $(r_o,\vartheta_o)$. We know for non-rotating black hole, the shape of the shadow is a perfect circle due to the spherically symmetrical system, and the rotation will lead to the shape deformation. In FIG.  \ref{fig:shadow a=0.1,0.5,0.95,change beta} and FIG.  \ref{fig:shadow a=0.95, theta=0.5pi, change beta and r_0}, we show the boundary of the shadow for the charged rotating black hole in conformal gravity.

The effects of parameter $\beta$ with different values of $a$ are shown in FIG.  \ref{fig:shadow a=0.1,0.5,0.95,change beta}. It is clear that the existence of $a$ and $\beta$ both enhances the deformation of shadow. This means  that the shadow of the charged rotating black hole in conformal gravity with  parameters $(a,\beta)$ and that of the Kerr black hole with a certain spin  may be coincident. While the influence of the charge parameter $\beta$ in conformal gravity on the shadow is qualitatively similar to that of the KN case \cite{Grenzebach:2014fha,Perlick:2021aok,Tsukamoto:2017fxq,Xavier:2020egv}.
In FIG.  \ref{fig:shadow a=0.95, theta=0.5pi, change beta and r_0}, we fix $a=0.95$ and $\beta=0.999 \beta_{ex}$. The left plot shows the influence of the viewing angle of the observer, which indicates that the shadow remains circular for a polar observer with $\vartheta_o=0$, while the shadow is maximally deformed for an observer in the equatorial plane with $\vartheta_o=\pi/2$. The right plot shows the influence of the distance between the observer and the black hole on the shadow, where the shadow is smaller
for the farther observer as expected.
\begin{figure} [H]
{\centering
\includegraphics[width=1.8in]{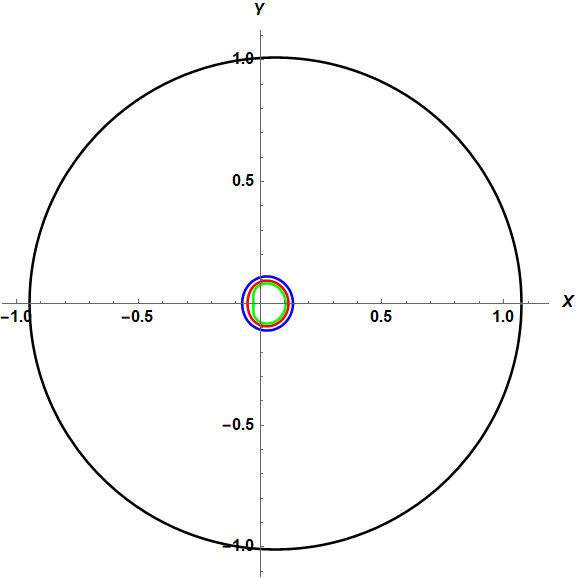}\hspace{0.5cm}
\includegraphics[width=1.8in]{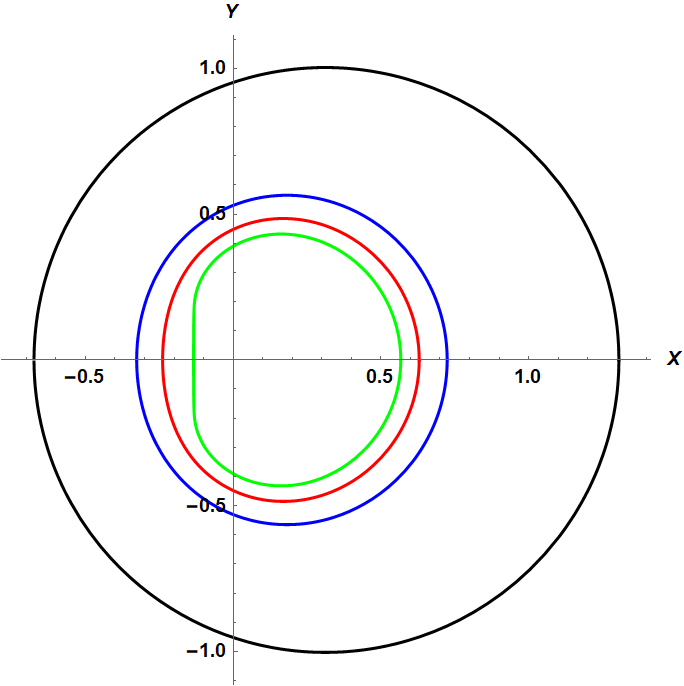}\hspace{0.5cm}
\includegraphics[width=1.8in]{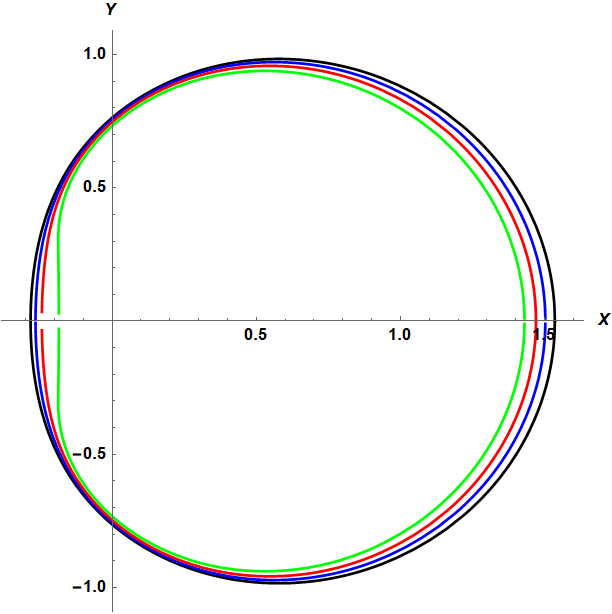}\hspace{0.5cm}
   \caption{Black hole shadows  seen by an equatorial observer $(\vartheta_{o}=\frac{\pi}{2})$ at $r_{o}=5$. Plots from left to right correspond to $a=0.1,a=0.5 $ and $a=0.95$. In each plot, the black, blue, red and green curves correspond to $\beta=(0,0.3,0.6,0.999)\beta_{ex}$.}   \label{fig:shadow a=0.1,0.5,0.95,change beta}}
\end{figure}
\begin{figure} [H]
{\centering
\includegraphics[width=2.3in]{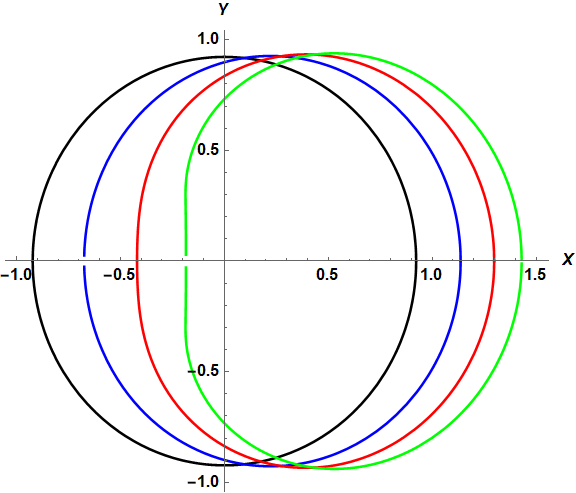}\hspace{0.5cm}
\includegraphics[width=2.0in]{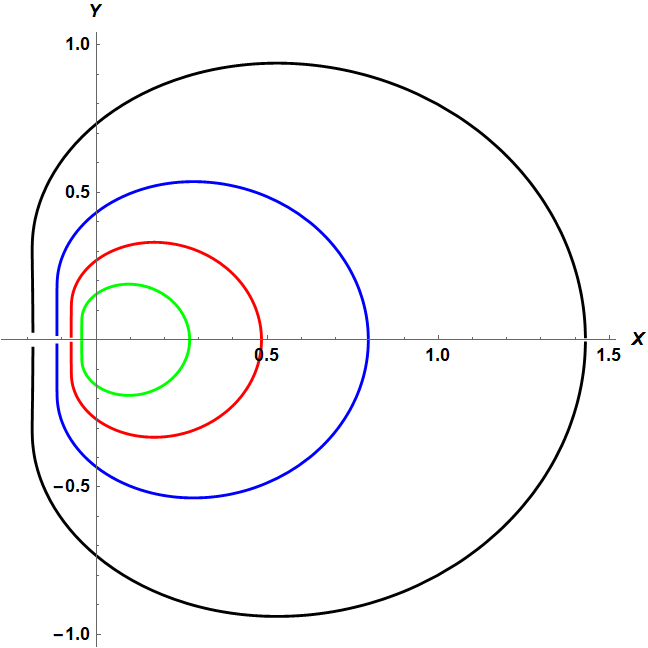}\hspace{0.5cm}
   \caption{LEFT: Black hole shadows with $r_o=5$ and different viewing angles. The shadow boundaries from left to right have  $\vartheta_o=0, \pi/8, \pi/4$, and $\pi/2$, respectively. RIGHT: Black hole shadows with $\vartheta_o=\pi/2$, and $r_o=5,10,20,50$ for boundaries from outer to inner. In both figures we have fixed $a=0.95$ and $\beta=0.999 \beta_{ex}$.}   \label{fig:shadow a=0.95, theta=0.5pi, change beta and r_0}}
\end{figure}

\section{Shadow observables and parameter estimation}\label{sec:parameter estimation}
To carefully study how the shadow observables are affected by the model parameters, we consider the black hole shadows observed at spatial infinity, i.e. $r_o\gg m$. In this case, as addressed in \cite{Perlick:2021aok}, the coordinates in \eqref{eq-Cartesian} can be transformed to $\bar\alpha=r_o X-a \sin\vartheta_o$ and $\bar\beta=r_o Y$, which are finally reduced as
\begin{eqnarray}
\bar\alpha{(r_p)}=-\frac{\xi{(r_p)}}{\sin \vartheta_{o}},~~~~
\bar\beta{(r_p)}=\pm\sqrt{\eta{(r_p)}+a^2 \cos^2\vartheta_{o}-\xi{(r_p)}^2\cot^2\vartheta_o}
\end{eqnarray}
where  $\xi(r_p)=L_E\mid_{r_p}$ and  $\eta=K_E-(L_E-a)^2\mid_{r_p}$.
Here $(\bar\alpha,\bar\beta)$ are the Bardeen's two impact parameters with length dimension describing the celestial sphere \cite{Cunningham:cgh}.

Subsequently, we show the boundary of shadow for the observer at spatial infinity in FIG. \ref{fig:shadow a=0.95, theta=0.5pi, change beta} and FIG.  \ref{fig:change beta, theta} in which the axes labels $(X,Y)$ represent $(\bar\alpha/m,\bar\beta/m)$.
We see that the boundary of black hole shadow closely depends on the parameters $a,\beta$ and $\vartheta_o$, and the tendencies are similar as that for the observer at finite distance.  It is noticed that the black hole parameters are expected to be associated and estimated from observations. Though the image of  M87* is mostly connected with Kerr black hole, the interesting point is if it is a black hole from MoG, the distortion of the shadow for a given spin parameter also arises due to the
presence of additional parameter as we show in our figures. 
Thus, instead of describing the similar properties, here we shall study how to estimate the parameters from the observables like the size and distortion of the black hole shadow.

\begin{figure} [H]
{\centering
\includegraphics[width=1.8in]{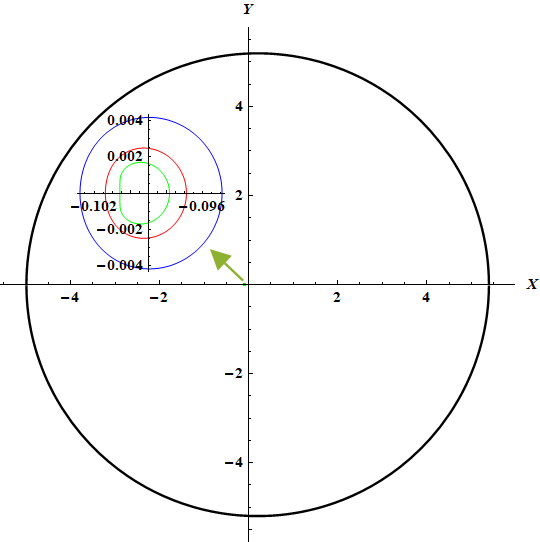}\hspace{0.5cm}
\includegraphics[width=1.8in]{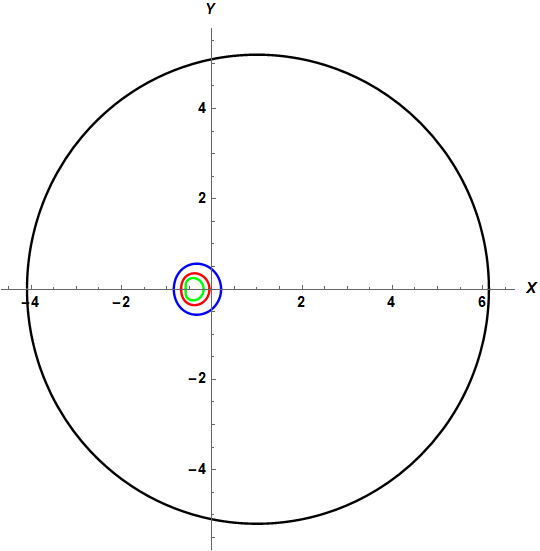}\hspace{0.5cm}
\includegraphics[width=1.9in]{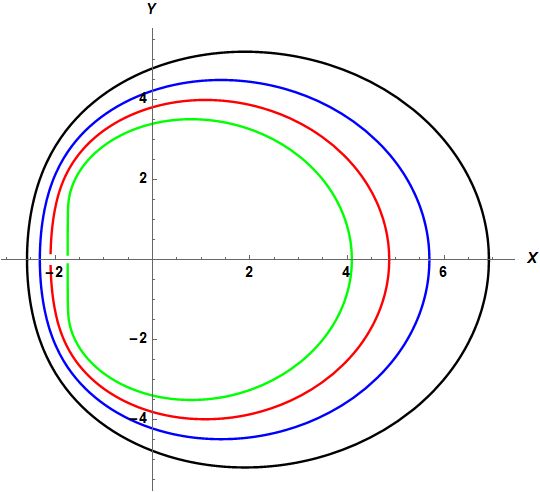}\hspace{0.5cm}
   \caption{Black hole shadow seen by an observer at infinity distance and $\vartheta_0=\pi/2$. We fix $a=0.1,a=0.5 $ and $a=0.95$ from left to right. In each plot, the black, blue, red and green curves correspond to $\beta=(0,0.3,0.6,0.999)\beta_{ex}$. }   \label{fig:shadow a=0.95, theta=0.5pi, change beta}}
\end{figure}
\begin{figure} [H]
{\centering
\includegraphics[width=2.0in]{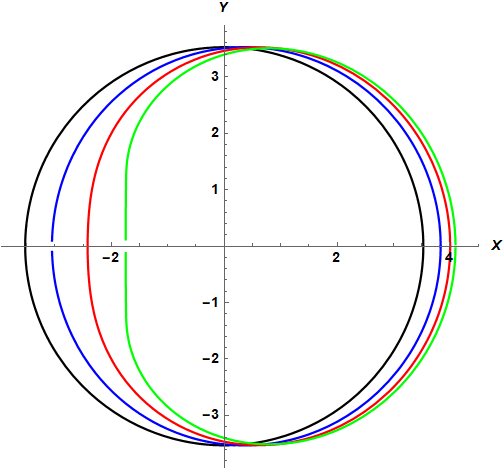}\hspace{0.5cm}
   \caption{Black hole shadows seen by an observer at infinity distance for different inclination angles: $\vartheta_o=0$ (black), $\pi/8$ (blue), $\pi/4$ (red), and~ $\pi/2$(green).
    We have fixed $a=0.95, \beta=0.999\beta_{ex}$. }   \label{fig:change beta, theta}}
\end{figure}

\subsection{Shadow size and deformation}\label{sec:Shadow size and deformation}
To describe the distortion and size of the charged rotating black hole in conformal gravity, we first study two characterized observables, $R_s$ and $\delta_s$ which were proposed by Hioki and Maeda \cite{Hioki:2009na}. Here $R_s$ is the radius of the reference circle for the distorted shadow and $\delta_s$ is the deviation of the left edge of the shadow from the reference circle boundary. For convenience, we denote the top, bottom, right and left of the reference circle as $(X_t, Y_t)$, $(X_b, Y_b)$, $(X_r,0)$ and $(X_l', 0)$, respectively and $(X_l, 0)$ as the leftmost edge of the shadow \cite{Ghosh:2020ece}. Subsequently, the definitions of the characterized observables are \cite{Hioki:2009na}
\begin{equation}\label{eq:Rs}
R_s=\frac{(X_t-X_r)^2+Y_t^2}{2\mid X_r-X_t\mid}, ~~~~~   \delta_s=\frac{\mid X_l-X_l'\mid}{R_s}.
\end{equation}

\begin{figure} [H]
{\centering
\includegraphics[width=2.5in]{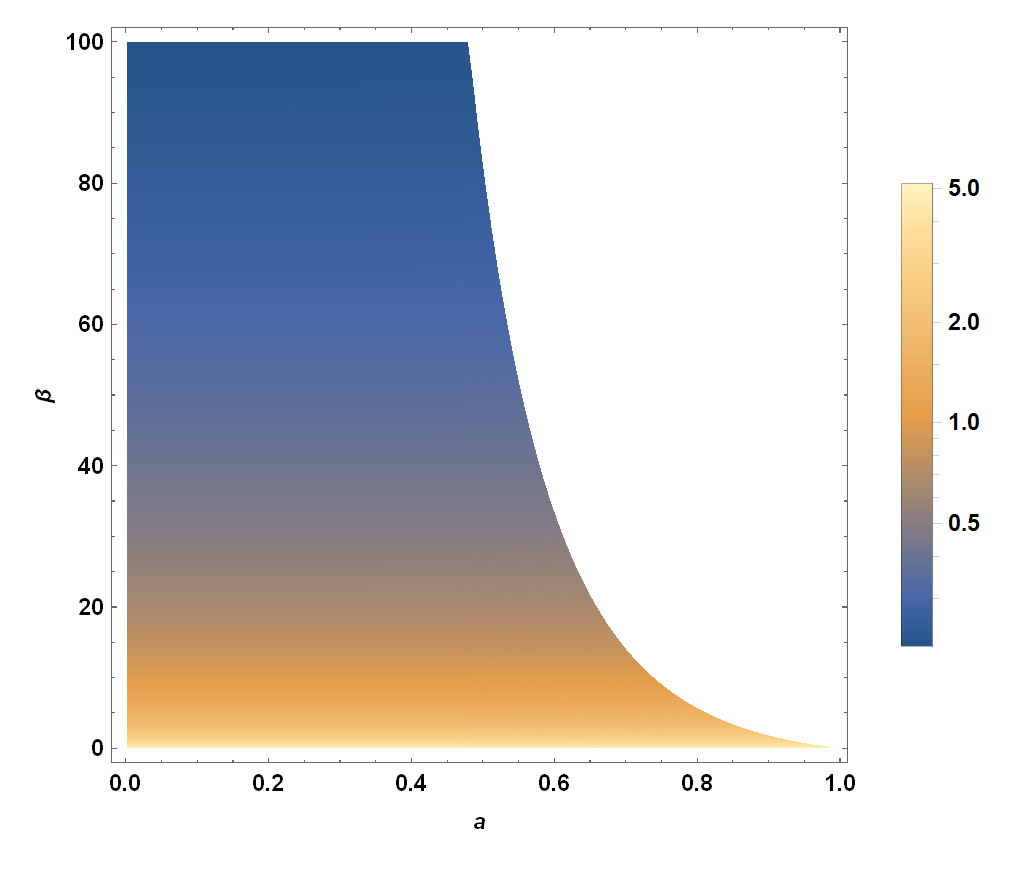}\hspace{1cm}
\includegraphics[width=2.5in]{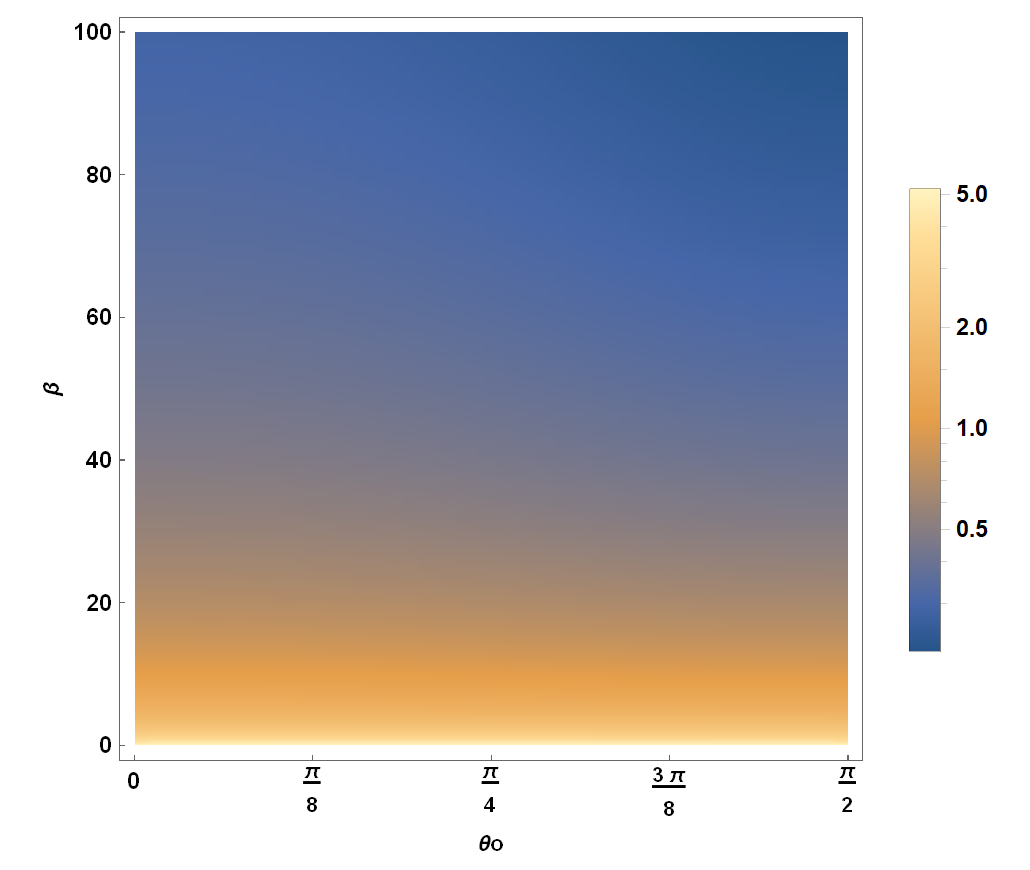}
   \caption{The density plots for the radius of the reference circle $R_s$ as a function of $a$ and $\beta$. Here we fix $\vartheta_o=\pi/2$ in the left plot  and $a=0.4$ in the right plot.}   \label{fig:Rs}}
\end{figure}

\begin{figure} [H]
{\centering
\includegraphics[width=2.5in]{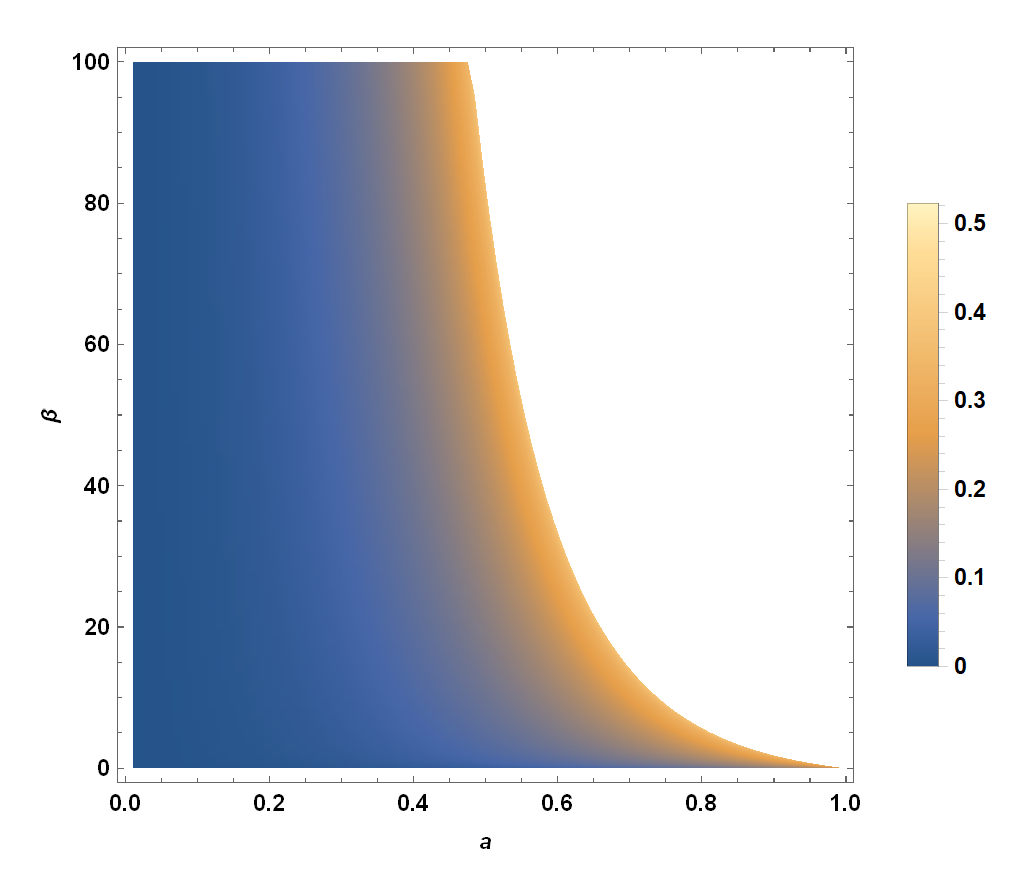}\hspace{1cm}
\includegraphics[width=2.5in]{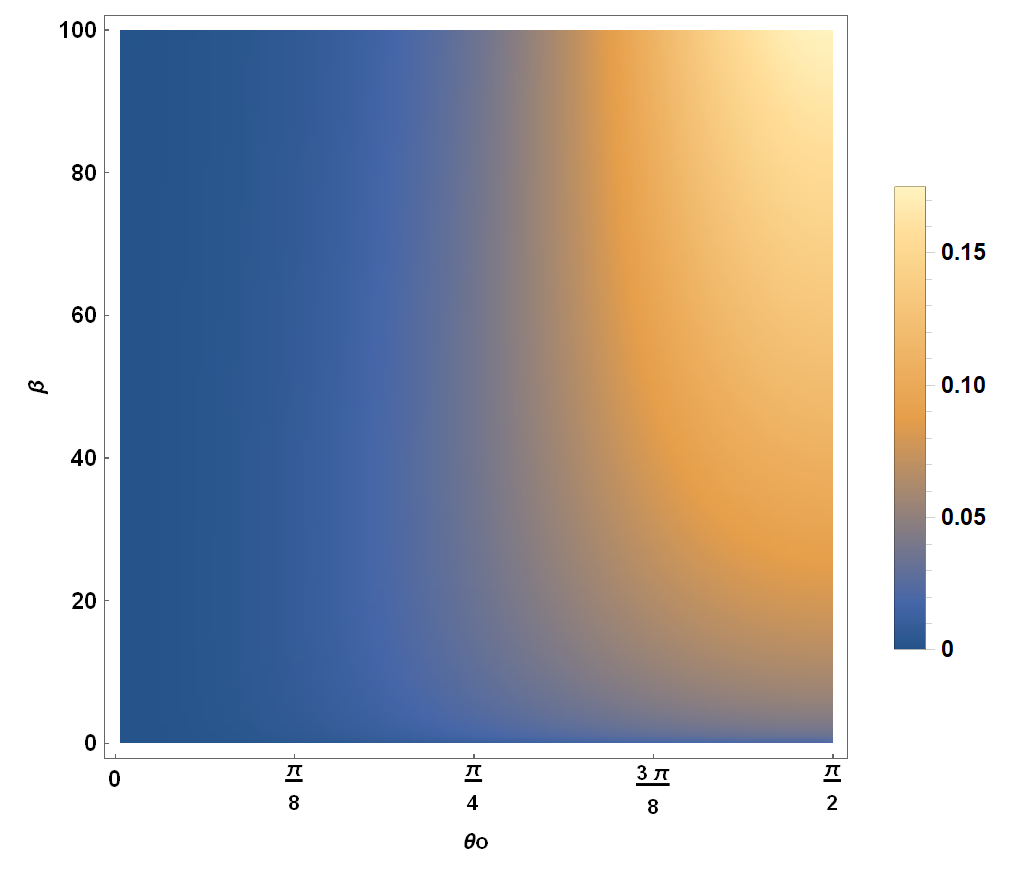}
   \caption{The density plot of the distortion $\delta_s$. Here we fix $\vartheta_o=\pi/2$ in the left plot  and $a=0.4$ in the right plot.}   \label{fig:deltaS}}
\end{figure}

\begin{figure} [H]
{\centering
\includegraphics[width=2.5in]{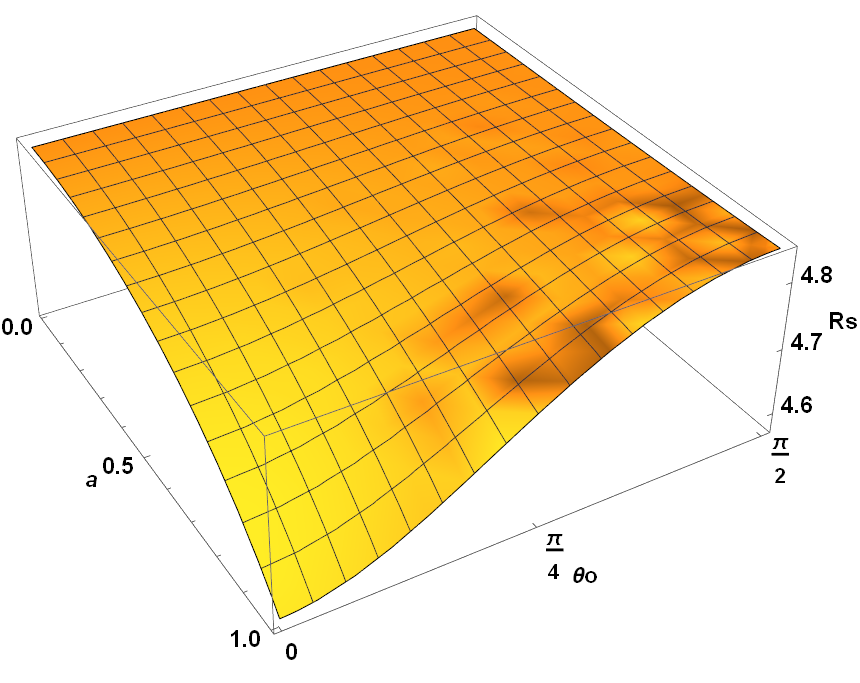}\hspace{1cm}
\includegraphics[width=2.5in]{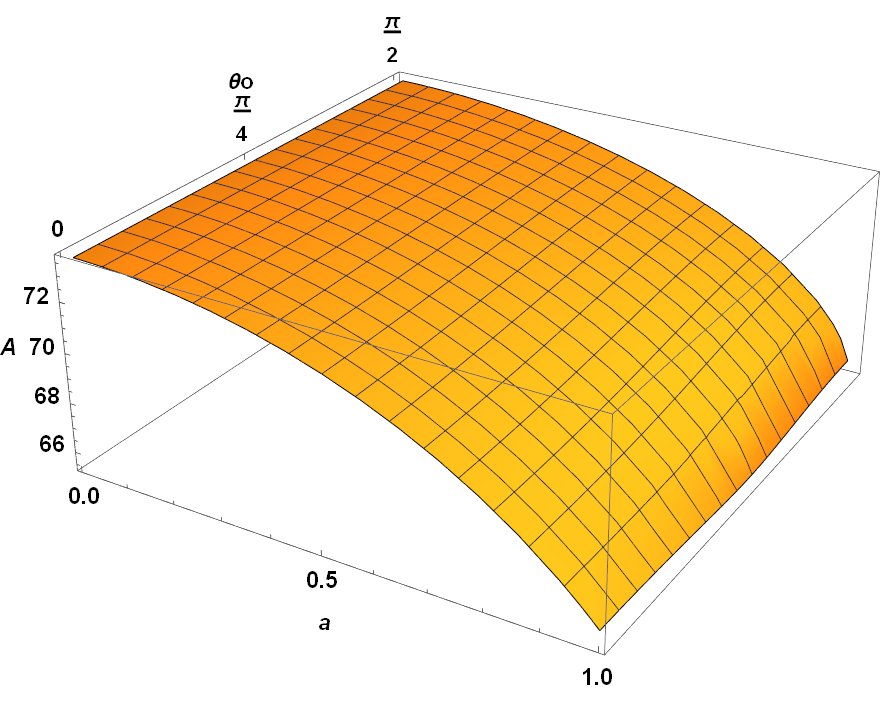}
   \caption{The 3D plots of radius $R_s$ (left) and area $A$ (right) of the black hole shadow. Here we fix $\beta=0.1$.}   \label{fig:RS-AA}}
\end{figure}

From the density plots of  $R_s$ and $\delta_s$ in FIG. \ref{fig:Rs} and FIG. \ref{fig:deltaS}, we see that the black hole parameters in conformal gravity have prints on the shadow size and shape.
FIG.  \ref{fig:Rs} shows that with the increase of the charge parameter $\beta$, the radius $R_s$ decreases rapidly. It is slightly affected by the spin paremeter $a$ and the inclination angle $\vartheta_o$, and their effects are enlarged in the left plot of FIG.  \ref{fig:RS-AA} from which we find that $R_s$ slightly decreases as $a$ increases while it increases as $\vartheta_o$ increases.  On the other hand, FIG.  \ref{fig:deltaS} shows that increasing $a$ or $\vartheta_o$, the distortion character $\delta_s$ increases which means that the shadow is more distorted as expected. Moreover, when $a$ or $\vartheta_o$ is small, the effect of $\beta$ on $\delta_s$ is slight but when they are large enough, $\beta$ has a profoundly incremental effect.
The above analysis further implies that comparing to Kerr black hole, the shadow radius of this charged rotating black hole in conformal gravity is always smaller but more distorted, which is similar to that of KN black hole \cite{Kumar:2019pjp}.

Since $R_s$ and $\delta_s$ may not accurately describe the shadow of some irregular black holes as they require the shadow of black holes to have certain symmetry. Then to characterize the shadow with any shape, Kumar and Ghosh proposed another two characterized observables,  the shadow area $A$ and oblateness $D$, which are defined as \cite{Kumar:2018ple}
\begin{eqnarray}
A=2\int Y{(r_p)}dX{(r_p)}=2\int^{r_p^{max}}_{r_p^{min}}\left(Y{(r_p)}\frac{dX{(r_p)}}{d{r_p}}\right)d{r_p},
~~~~~
D=\frac{X_r-X_l}{Y_t-Y_b}.
\label{eq-A}
\end{eqnarray}

\begin{figure} [H]
{\centering
\includegraphics[width=2.5in]{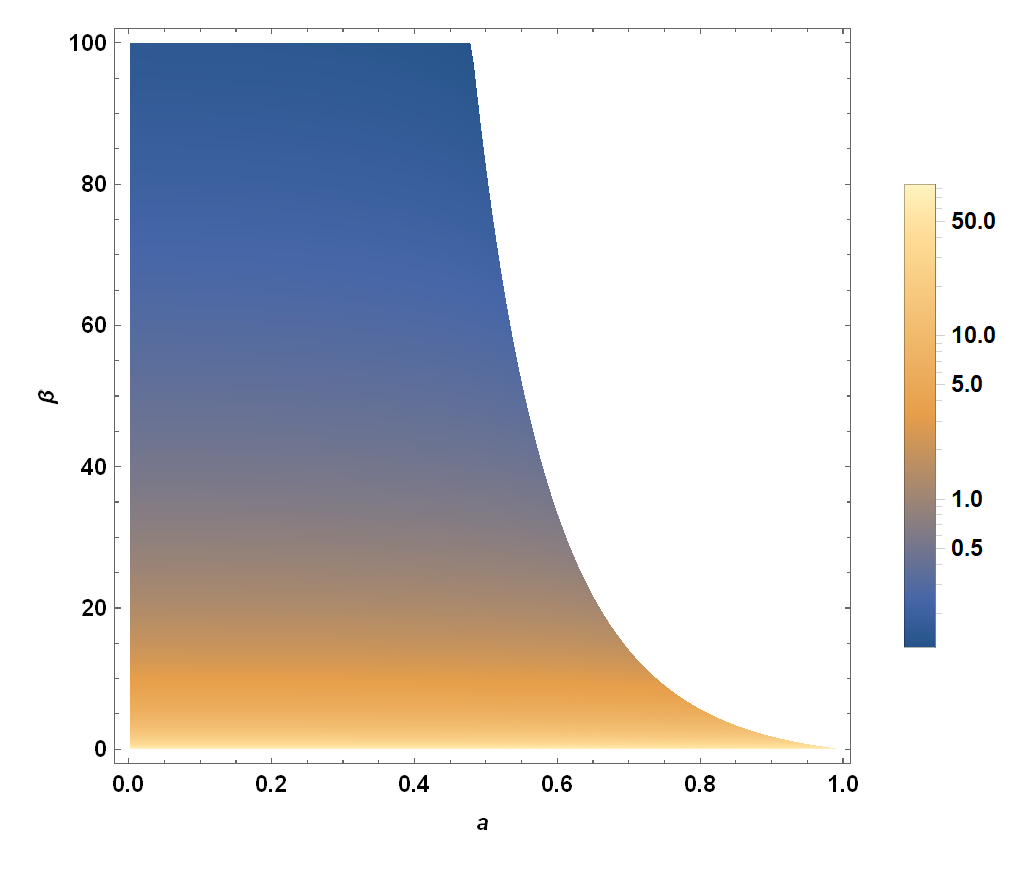}\hspace{1cm}
\includegraphics[width=2.5in]{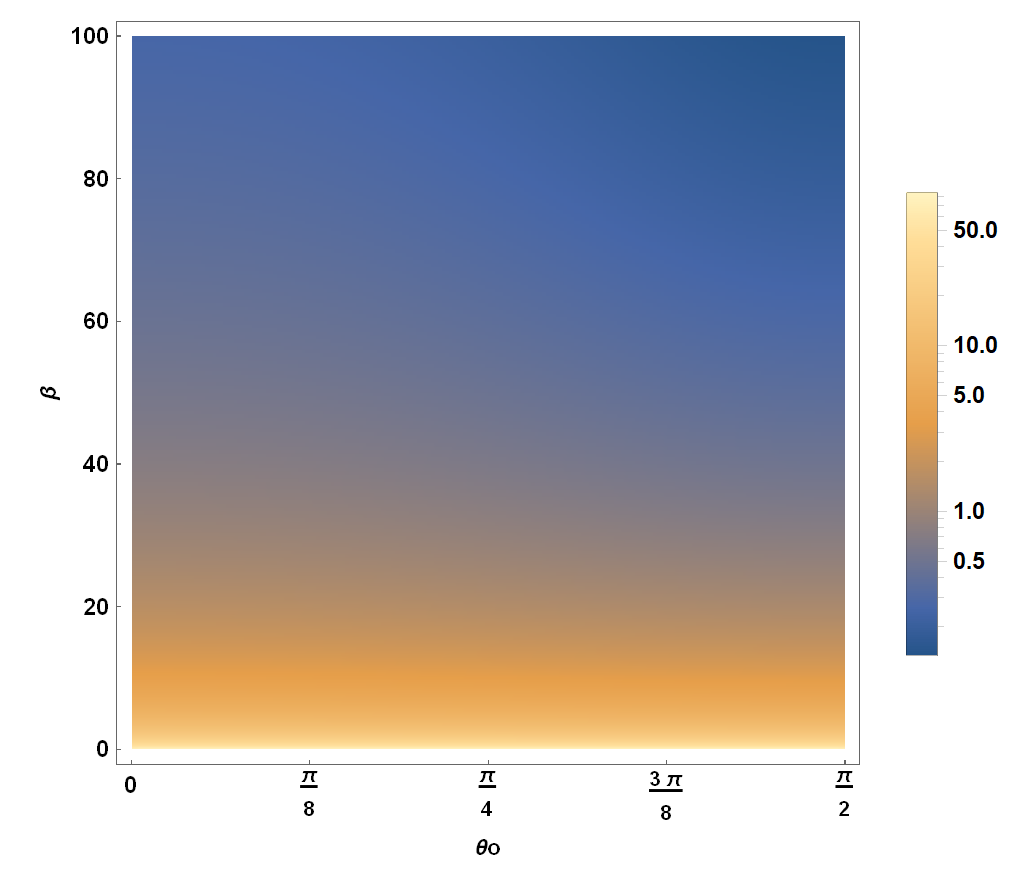}
   \caption{The density plots of the shadow area $A$. Here we fix $\vartheta_o=\pi/2$ in the left plot  and $a=0.4$ in the right plot.}   \label{fig:A}}
\end{figure}

\begin{figure} [H]
{\centering
\includegraphics[width=2.5in]{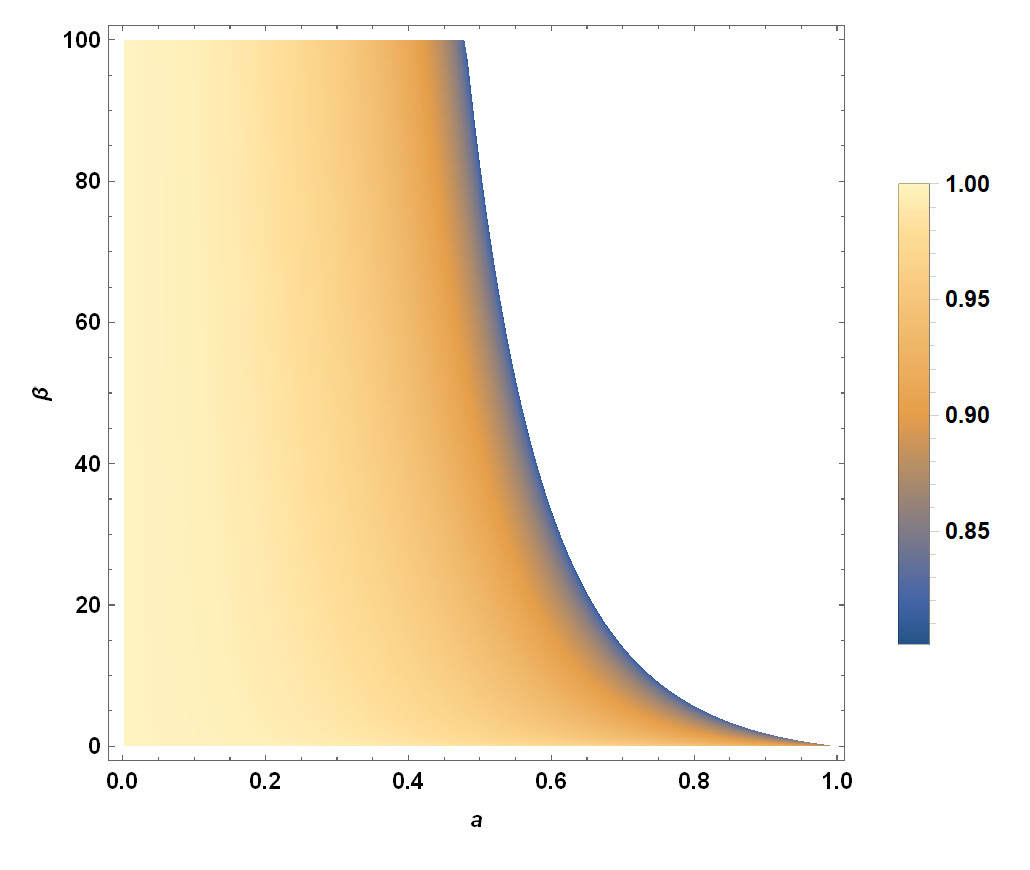}\hspace{1cm}
\includegraphics[width=2.5in]{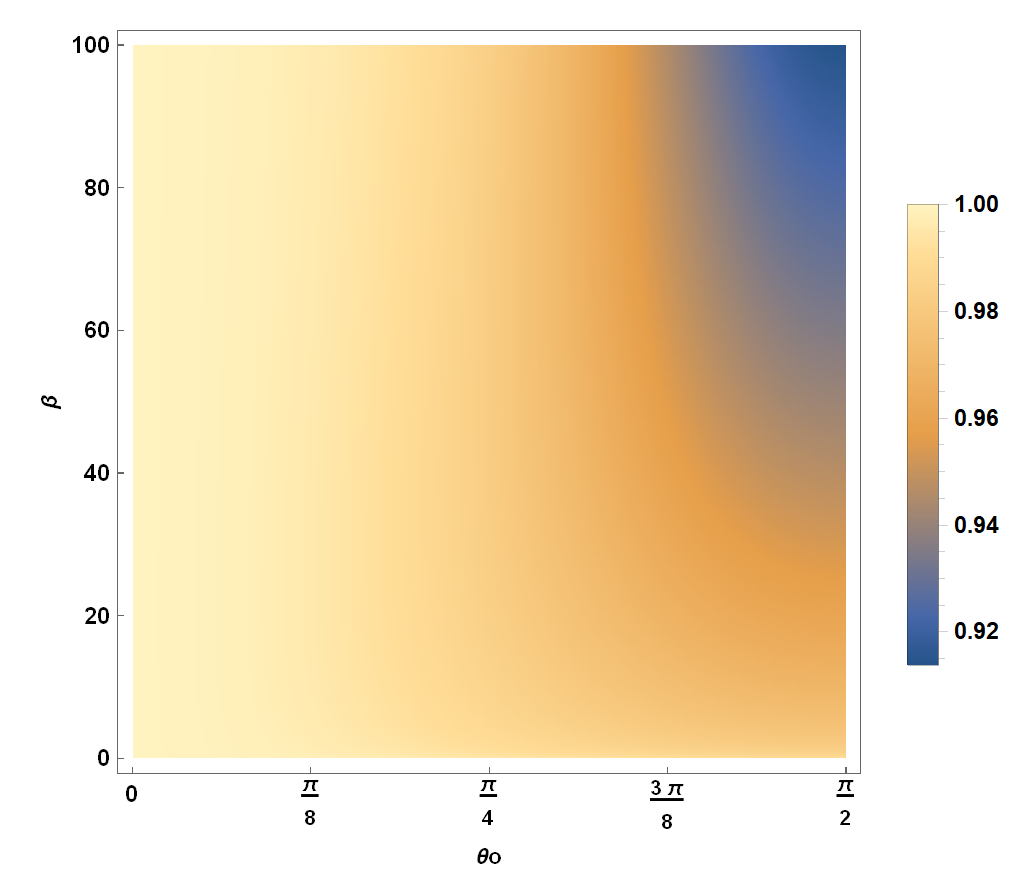}
   \caption{The density plots of the oblateness $D$. Here we fix $\vartheta_o=\pi/2$ in the left plot  and $a=0.4$ in the right plot.}   \label{fig:D}}
\end{figure}

It was found in \cite{Tsupko:2017rdo} that $D=1$ for Schwarzschild black hole and $\sqrt{3}/{2} \le D<1$ for Kerr black hole in the view of an equatorial observer, where  $D=\sqrt{3}/{2}$ is for the extremal case.
In FIG. \ref{fig:A}-\ref{fig:D}, we show the density plots of $A$ and $D$ for the shadow of the charged rotating black hole in conformal gravity. The area $A$ monotonously decreases as $\beta$ increases. The influence of $a$ and $\vartheta_o$ is enlarged in the right plot of FIG.  \ref{fig:RS-AA}, which shows that the area slightly decreases as the spin increases while the effect of $\vartheta_o$ is negligible.
As $\beta$ increases, the oblateness $D$ becomes smaller which is significant  near the extremal case. In addition, as $a$ or $\vartheta_o$ increases with the other fixed, $D$ also has decremental tendency. The above analysis also implies that the shadow of the charged rotating black hole in conformal gravity is smaller and more distorted than that of Kerr black hole, which matches our aforementioned finding.

\begin{figure} [H]
{\centering
\includegraphics[width=2.in]{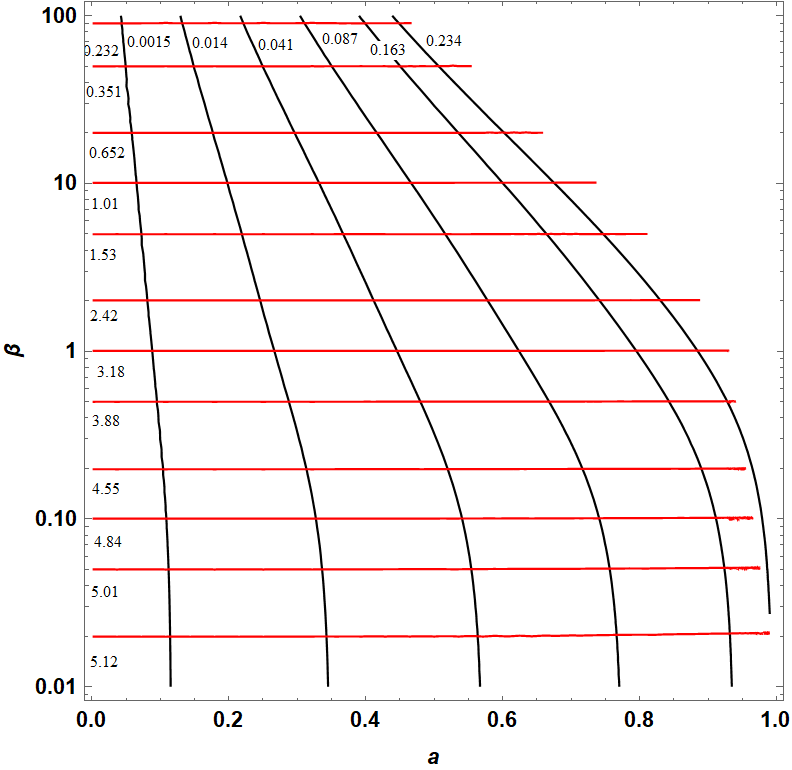}\hspace{1cm}
\includegraphics[width=2.in]{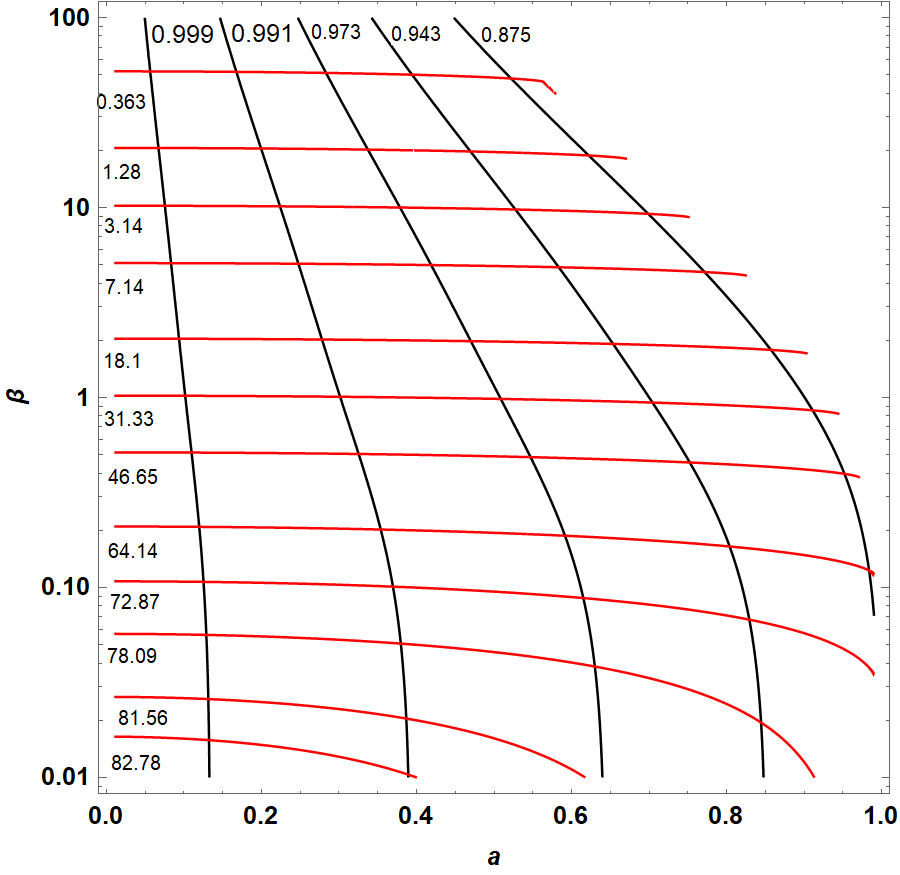}
   \caption{LEFT: the contour plot for shadow observables $R_s$ (red) and $\delta_s$ (black) in the parameter plane $(a,\beta)$ of the charged rotating black hole in conformal gravity. RIGHT: the contour plot for shadow observables area $A $ (red) and oblateness
   $D$ (black).}   \label{fig:contourPlot}}
\end{figure}

So far, we have explored how the black hole parameters leave prints on the two couples of shadow observables, i.e. $(R_s,\delta_s)$ and  $(A,D)$. Then with given values of $(R_s,\delta_s)$ or $(A,D)$, we can find their contour intersection in the parameters $a-\beta$ plane to estimate the parameters of the charged rotating black hole in conformal gravity. This method of black hole parameter estimation  from its shadow observables
has been implemented in \cite{Hioki:2009na,Kumar:2018ple,Afrin:2021imp,Afrin:2021wlj}. Here, we fix $\vartheta_o=\pi/2$ and show the contour plots of $R_s$ and $\delta_s$ as well as $A$ and $D$ in FIG.  \ref{fig:contourPlot}  in which the intersection point of $R_s (A)$ and $\delta_s (D)$  uniquely determines the black hole parameters $a$ and $\beta$.

\subsection{Energy emission rate}
Apart from being used to estimate the model parameters, the shadow observables are also helpful to predict various interesting astronomical
phenomena \cite{Kumar:2018ple,Kumar:2020owy,Afrin:2021imp}. In this subsection, we shall analyze the energy emission rate for the charged rotating black hole in conformal gravity using the shadow observables.
For an observer at infinity distance, the shadow of a spherically symmetric black hole coincides to a high energy absorption cross section, which oscillates around a constant limiting value $\delta_{lim}$. It was addressed in \cite{Wei:2013kza} that $\delta_{lim}$ is connected with the black hole shadow via
\begin{equation}
\delta_{lim}\approx \pi R_s^2
\end{equation}
with $R_s$ defined in \eqref{eq:Rs}, hence
the energy emission rate for a rotating black hole can be calculated as
\begin{equation}
\frac{d^2 E(\varpi)}{d \varpi dt}=\frac{2 \pi^2 R_s^2}{e^{\varpi/T}-1}\varpi^3
\end{equation}
where $\varpi$ is the photon frequency and $T$ is the Hawking temperature at the event horizon of the black hole.

The energy emission rate in this proposal has been widely studied in GR and MoG.  Now we intend to discuss the energy emission rate for the charged rotating black hole \eqref{eq-metric}, the Hawking temperature of which is
\begin{equation}
T=\frac{-3a^4+r_+^4(3+4\beta)+\sqrt{3}B(r_+^2-a^2)}{4\pi r_+^2(a^2+r_+^2)(3a^2+3r_+^2+\sqrt{3}B)},
\end{equation}
with $B=3a^4+6a^2 r_+^2+r_+^4(3+4\beta)$.

\begin{figure} [H]
{\centering
\includegraphics[width=1.8in]{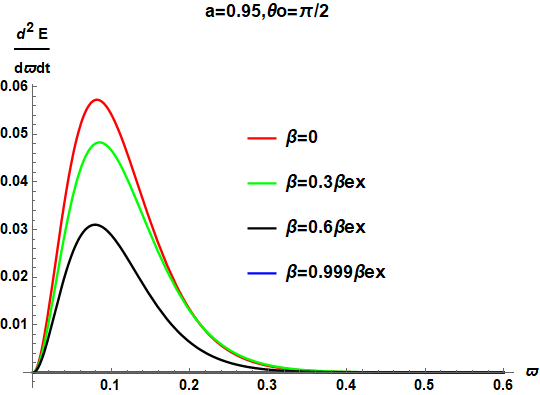}\hspace{0.5cm}
\includegraphics[width=1.8in]{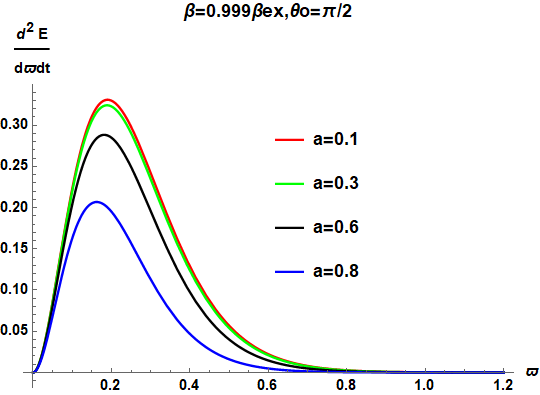}\hspace{0.5cm}
\includegraphics[width=1.8in]{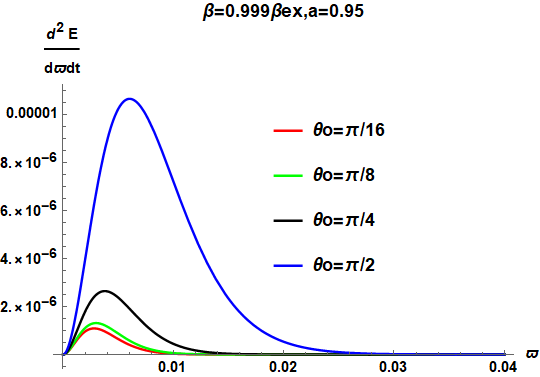}\hspace{0.5cm}
   \caption{The distribution of the energy emission rate in terms of the photon frequency $\varpi$ with various values of parameters $\beta$, $\vartheta_o$ and $a$.}   \label{fig:energy-emission-rate}}
\end{figure}

In FIG.  \ref{fig:energy-emission-rate}, we present the behavior of the energy emission rate as a function of photon frequency. The left and middle plots show that the peak of the emission rate decreases as both $\beta$ and $a$ increases and the peak shifts to lower frequency, while the right plot shows that the inclination angle has the opposite effect on the emission rate.

\section{Constraints from EHT observations of M87*}\label{sec:constraints}

The black hole image of M87*  photographed by the EHT is { crescent shaped}, and its deviation from circularity in terms of the root-mean-square distance from the average radius of the shadow  is $\Delta C \lesssim 0.1$. The axis ratio is $1<D_x\lesssim 4/3$ while the angular diameter is $\theta_d=42\pm3 \mu a s$ \cite{EventHorizonTelescope:2019dse,EventHorizonTelescope:2019ths,EventHorizonTelescope:2019pgp}. The preliminary analysis of the image of M87* by EHT collaboration refers to the Kerr black hole whose parameters are constrained by the above observations, but the results can not rule out the alternative black holes in GR or the rotating black holes in MoG. Thus, the shadow observables $\Delta C$, $D_x$ and $\theta_d$ could also be used to constrain the parameters of black holes in MoGs, and some attempts can be seen in
\cite{Cunha:2019ikd,EventHorizonTelescope:2021dqv,Khodadi:2020jij,Bambi:2019tjh,
Afrin:2021imp,Kumar:2019pjp,Ghosh:2020spb,Afrin:2021wlj,Jha:2021bue,Khodadi:2021gbc}.

In this section, we presuppose the M87* a rotating charged  black hole in conformal gravity and will use the EHT observations to constrain the parameters $a$ and $\beta$. To this end, we shall first review the definition of
$\Delta C$, $D_x$ and $\theta_d$, and show their density plots in the parameter space $(a,\beta)$.

To describe the circularity deviation $\Delta C$, we have to recall from subsection
\ref{sec:Shadow size and deformation} that the distorted black hole shadow is always compared with a reference circle. The geometric center of the shadow $(X_c, Y_c)$ is connected with the edges of the  shaped boundary via $({X_c=\frac{X_r+X_l}{2}}, Y_c=0)$, and with this point as the origin, the boundary of a black hole shadow can be described by the polar coordinates $(\phi,R(\phi))$ where
\begin{equation}
\begin{split}
\phi=\tan^{-1}\left(\frac{Y-Y_C}{X-X_c}\right),~~~~
R(\phi)=\sqrt{(X-X_c)^2+(Y-Y_c)^2},
\end{split}
\end{equation}
while the average radius of the shadow is
\begin{equation}
\bar{R}=\frac{1}{2\pi}\int^{2\pi}_{0} R(\phi)d\phi.
\end{equation}
Then the circularity deviation $\Delta C$ which measures the deviation from a perfect circle is defined by \cite{Afrin:2021imp}
\begin{equation}\label{eq-DeltaC}
{\Delta C=\frac{1}{\bar{R}}\sqrt{\frac{1}{2\pi}\int^{2\pi}_{0}(R(\phi)-\bar{R})^2 d\phi}.}
\end{equation}

The axis ratio is given by \cite{Banerjee:2019nnj}
\begin{equation}\label{eq-Dx}
D_x=\frac{1}{D}=\frac{Y_t-Y_b}{X_r-X_l},
\end{equation}
where the  oblateness $D$ has been defined in \eqref{eq-A}. In fact, $D_x$ could be seen as another way of defining the circular derivation since the emission ring reconstructed in EHT images is close to circular with an axial ratio of $4:3$, which indeed also correspond to $\Delta C\lesssim 0.1$ \cite{EventHorizonTelescope:2019dse}.

Another observable from the EHT collaboration is the angular diameter of the shadow which is defined as \cite{Kumar:2020owy}
\begin{equation}\label{eq-theta-d}
\theta_d=2\frac{R_a}{d}\,
\end{equation}
where $R_a=\sqrt{\frac{A}{\pi}}$ with $A$ defined in \eqref{eq-A} is known as the shadow areal radius and $d$ is the distance of the M87* from the earth.

It is obvious from the formulas \eqref{eq-DeltaC}, \eqref{eq-Dx} and \eqref{eq-theta-d} that  $\Delta C$, $D_x$ and $\theta_d$ depend on the black hole parameters. Assuming M87* the current charged rotating black hole in conformal gravity, we could evaluate them for the metric \eqref{eq-metric} and use the EHT observations $\Delta C \lesssim 0.1$, $D_x\in(1,4/3]$ and $\theta_d\in[39,45] \mu a s$ to give constraints on the parameters $a$ and $\beta$.  In addition, we know that the shadow is maximally deformed at large inclination angle $\vartheta_o=\pi/2=90^{\circ}$, while the inclination angle (with respect to the line of sight)
is estimated to be $\vartheta_o=17^{\circ}$ in the  M87* image if considering the orientation of the relativistic
jets \cite{CraigWalker:2018vam}. So we shall then show our computational results for both $\vartheta_o=90^{\circ}$ and $\vartheta_o=17^{\circ}$.

\begin{figure} [H]
{\centering
\includegraphics[width=2.5in]{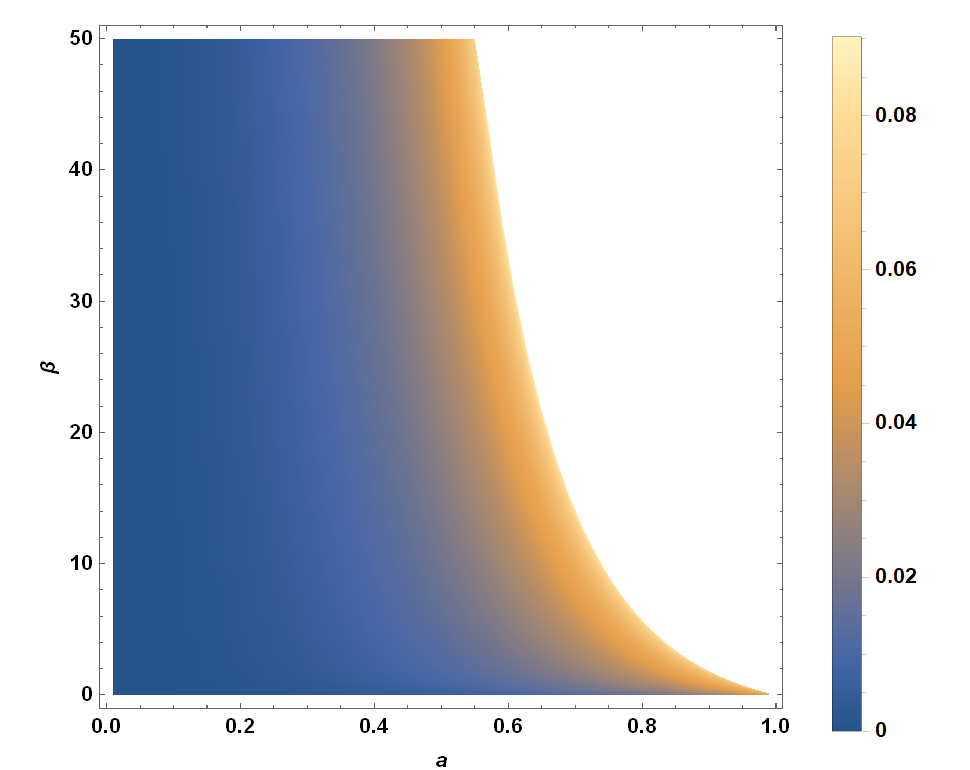}\hspace{1cm}
\includegraphics[width=2.5in]{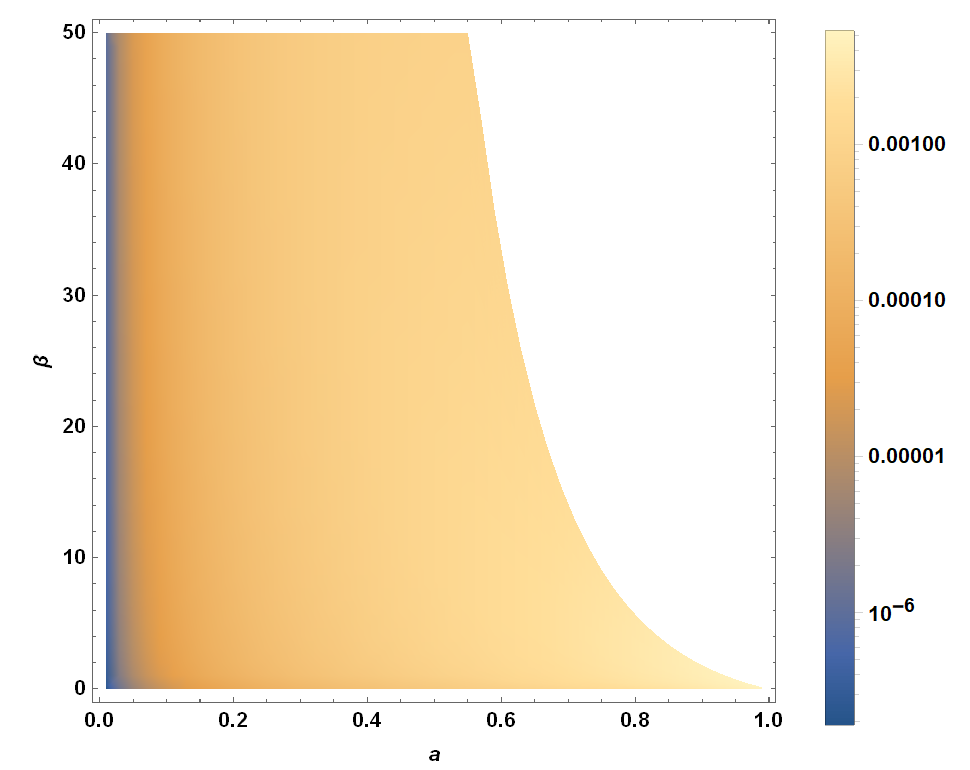}
   \caption{The density plots of the circularity deviation $\Delta C$. The left plot is for $\vartheta_o=90^{\circ}$ while the right plot is for $\vartheta_o=17^{\circ}$.}   \label{fig:deltaC}}
\end{figure}

\begin{figure} [H]
{\centering
\includegraphics[width=2.5in]{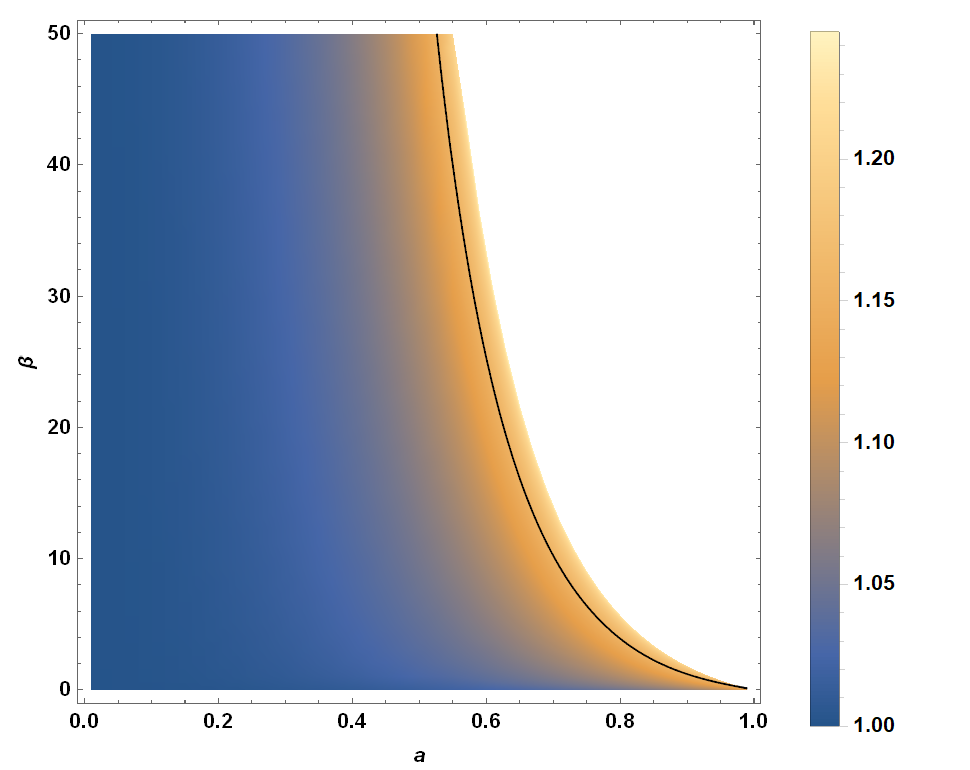}\hspace{1cm}
\includegraphics[width=2.5in]{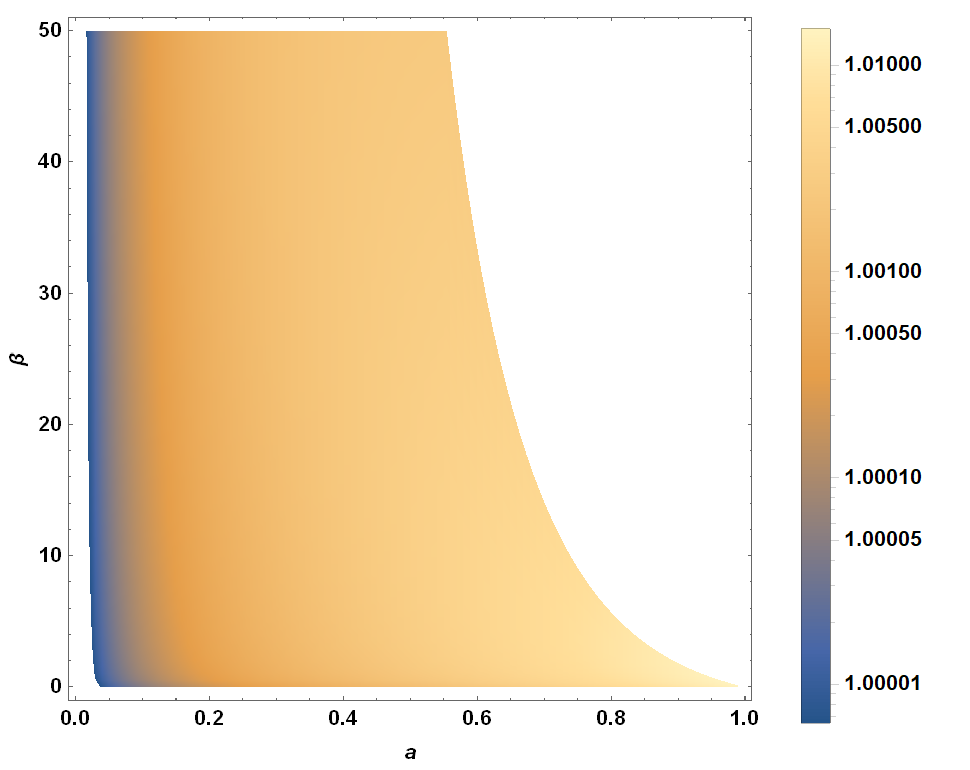}
   \caption{The density plots of the axial ratio $D_x$.  The left plot is for $\vartheta_o=90^{\circ}$ while the right plot is for $\vartheta_o=17^{\circ}$.  The black curve in the left plot denotes $D_x=2\sqrt{3}/3$ contour which is the upper bound for Kerr black hole.}  \label{fig-Dx}}
\end{figure}

We give the density plots of the circularity deviation $\Delta C$ in FIG.  \ref{fig:deltaC}, which shows that the shadows of the charged rotating black hole in conformal gravity satisfy $\Delta C \lesssim 0.1$ for all theoretically allowed parameters. Moreover, we also show the density plots of $D_x$ in FIG.  \ref{fig-Dx}. We see that for the entire parameter space, the axial ratio is within the observation constraint $D_x\in(1,4/3]$, which is consistent with the conclusion from $\Delta C \lesssim 0.1$ as we expect. In addition, in order to compare with $D_x\in(1,2\sqrt{3}/3]$ in Kerr black hole, we tend to show the contour with  $D_x=2\sqrt{3}/3$ in the calculation. For $\vartheta=90^{\circ}$, we see that in the current background, though all parameters satisfy $D_x<4/3$, their is still some parameter space with $D_x>2\sqrt{3}/3$. It means that if in the future, the EHT experiment is improved,  the observation $2\sqrt{3}/3<D_x<4/3$ even could rule out Kerr black hole in the center, and the current charged rotating black hole in conformal gravity could be a candidate. Nevertheless, for $\vartheta=17^{\circ}$, all the parameters give $1<D_x\leq2\sqrt{3}/3$, so one cannot distinguish GR and the conformal gravity.


\begin{figure} [H]
{\centering
\includegraphics[width=2.5in]{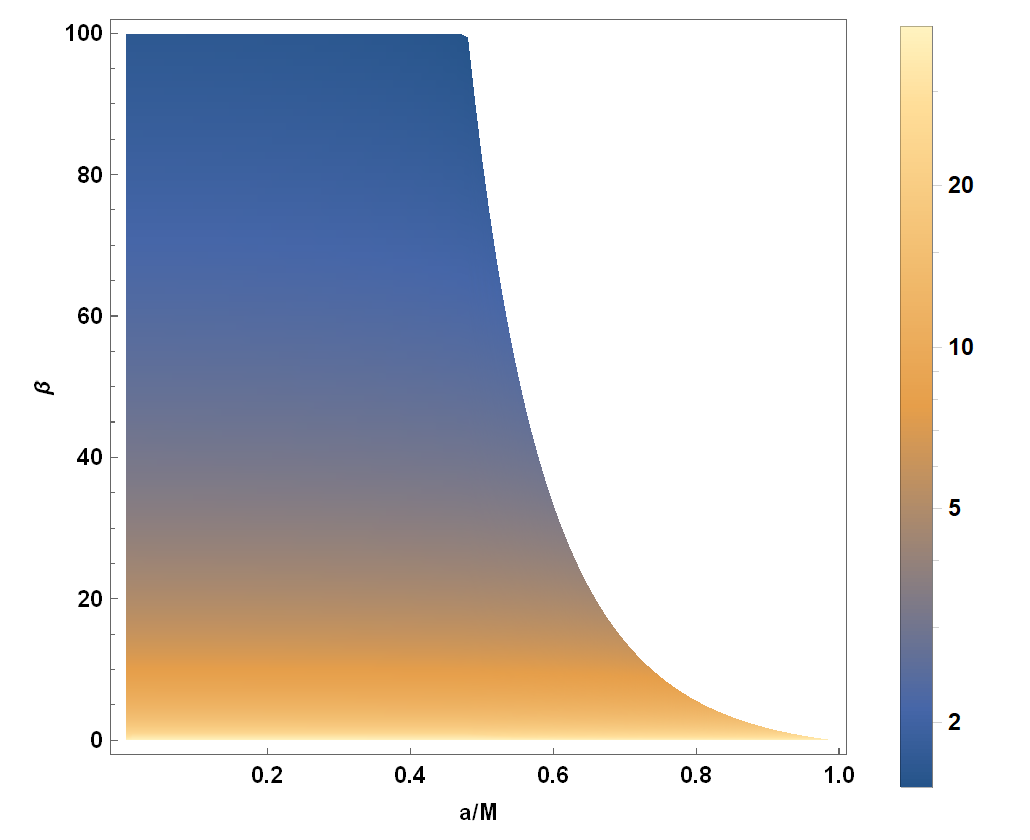}\hspace{1cm}
\includegraphics[width=2.5in]{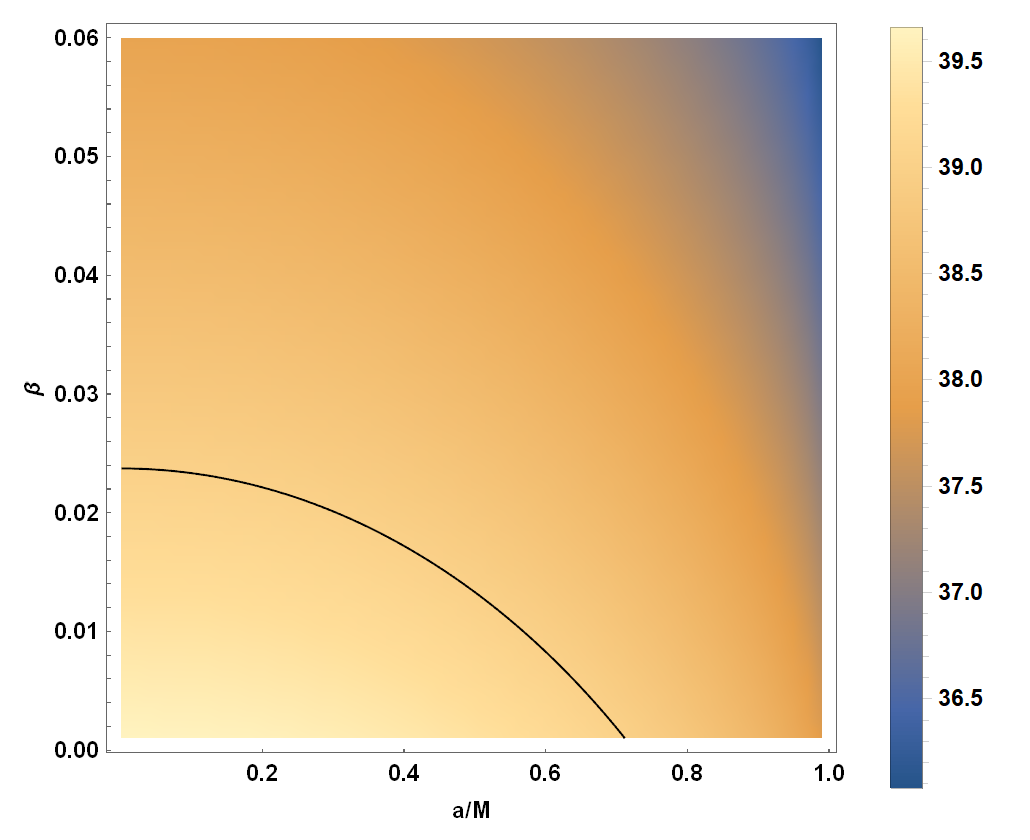}
\includegraphics[width=2.5in]{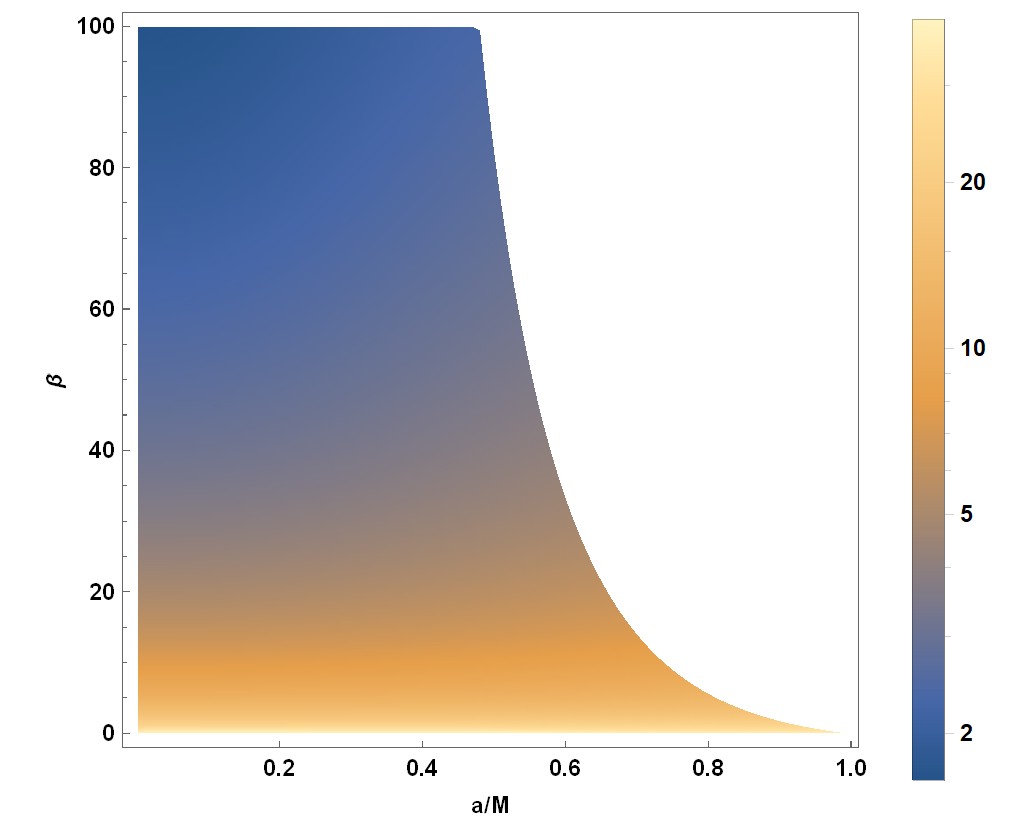}\hspace{1cm}
\includegraphics[width=2.5in]{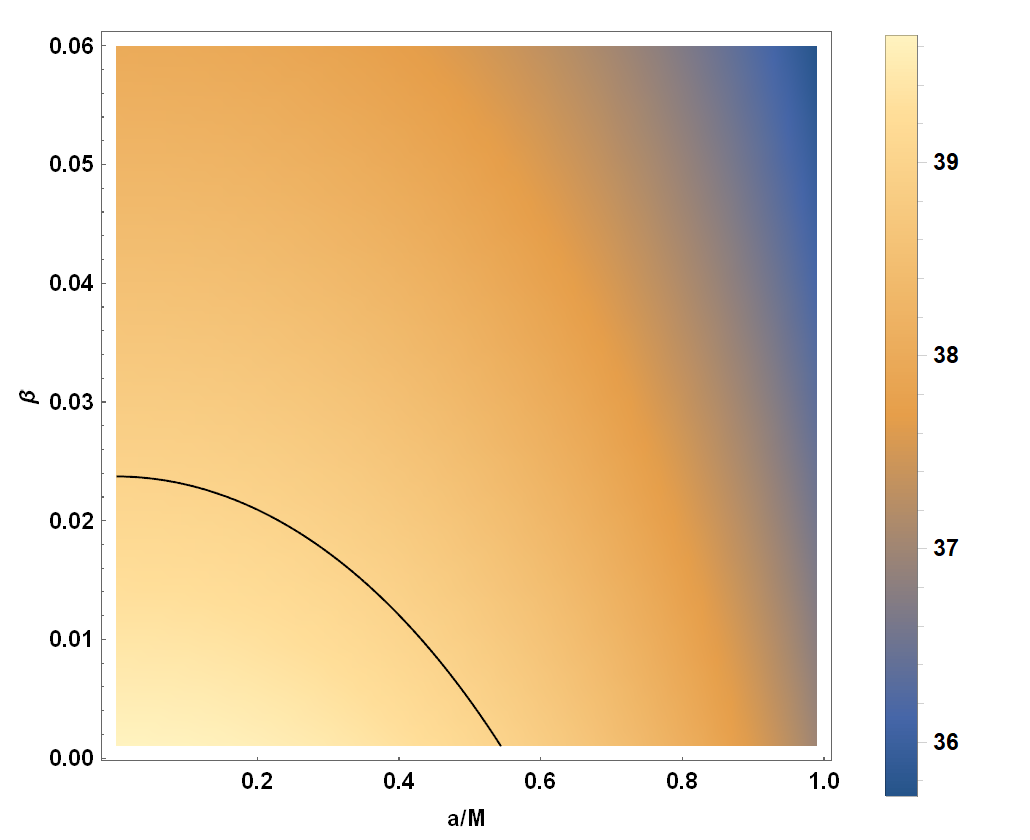}
   \caption{The density plots of the angular diameter $\theta_d$. The upper panels are for $\vartheta_o=90^{\circ}$ while the bottom panels are for $\vartheta_o=17^{\circ}$.
   The black curves in the right enlarged plots correspond to $\theta_d=39~\mu a s$.}   \label{fig:thetaD}}
\end{figure}

In FIG.  \ref{fig:thetaD}, we present the density plots of $\theta_d$ for the charged rotating black hole in conformal gravity. In the calculation, we set $d=16.8 Mpc$ and the black hole mass as $m=6.5\times 10^9 M_\odot$ as estimated by EHT collaboration. The enlarged plots in the right panel clearly show that only the parameter space at the left corner enclosed by the $\theta_d=39 \mu a s$ contour (the black curve) is consistent with the EHT observations of M87*, indicating that $\theta_d$ gives upper limit on both $a$ and $\beta$ in the charged rotating black hole in conformal gravity \eqref{eq-metric}.
Moreover, it is not difficult to find that the constraint on $a$ at $\vartheta=17^{\circ}$ is stricter  than that at $\vartheta=90^{\circ}$, but the difference of their effects on $\beta$ is slight.

\section{Closing remarks}
The published EHT observations on black hole image are consistent with those for Kerr black hole predicted by GR, but the current experimental outcome can not rule out alternatives to the Kerr black hole as well as other theories of gravity. In this paper, we considered a charged rotating black hole in conformal gravity which has remarkable implements  in cosmological  and holographical framework. The charged related term in the current black hole has different falloff from that in KN black hole, such that it exhibits different configurations. The charge parameter $\beta$ would decrease the size of both Cauchy and event horizons, of which the tendency is similar to that in KN black hole but with a different slope. Also, the size of event horizon in extremal case decreases as the charge parameter increases, in contrast to the independent situation in KN black hole. Moreover, the falloff term also has influence on the static limit surfaces, ergoregions, the causality violating regions and photon regions as we explicitly presented in FIG. \ref{fig:photon region a=0.5}-\ref{fig:photon region a=0.95}.

Then we figure out the shadow boundary of the black hole with various cases for observers at both finite and infinity distances. The effects of the spin parameter, the charge parameter, the inclination angle and the distance on the shadow shape can be clearly seen in FIG. \ref{fig:shadow a=0.1,0.5,0.95,change beta}-\ref{fig:change beta, theta}, which
are qualitatively similar to that of Kerr or KN black hole \cite{Perlick:2004tq,Hasse:2005vu,Grenzebach:2014fha}. Then focusing on the shadow cast for observer at infinity,  we systematically analyze the shadow observables that characterize the shadow size and shape, namely shadow radius $R_s$, distortion $\delta_s$, the shadow area $A$ and oblateness $D$. It was found that comparing with Kerr black hole, the black hole shadow is smaller and more distorted with the increasing of the charge parameter. Our analysis also indicates that the shadow observables could be used to estimate the parameters $(a,\beta)$ of the charged rotating black hole in conformal gravity.

Finally, we considered  the M87* in EHT experiment as the current charged rotating black hole in conformal gravity, and used the EHT constraints on the circularity deviation $\Delta C$, the axial ratio $D_x$ and the angular diameter $\theta_d$ to constrain the black hole parameters. For inclination angles $\vartheta_0=90^\circ$ and $\vartheta_0=17^\circ$,  the entire $(a,\beta)$ space satisfies $\Delta C\lesssim 0.1$ and $1<D_x\lesssim 4/3$.  It is worthwhile to point out that for $\vartheta=90^{\circ}$, some parameter space would give $2\sqrt{3}/3<D_x<4/3$ where $D_x=2\sqrt{3}/3$ is the upper bound for Kerr black hole. While  for $\vartheta=17^{\circ}$, all the parameters give $1<D_x<2\sqrt{3}/3$, so one cannot distinguish GR and the conformal gravity in this case. The $39~\mu a s\le \theta_d\le 45~\mu a s$ gives upper bounds
on both $a$ and $\beta$ and constrain the parameter space into a small portion.
To conclude, in plenty of parameter points $(a,\beta)$, the charged rotating black hole shadows are consistent with that in EHT observations of M87*. Our findings indicate that the charged rotating black hole in conformal gravitys with those parameters could be candidates for astrophysical black holes. Moreover, for the equatorial observer, the constraint of EHT on the axial ratio $D_x$ could help us to distinguish Kerr black hole and the current charged rotating black hole in conformal gravity in some parameter space.

\begin{acknowledgments}
We appreciate Xi-Jing Wang for helpful discussion. This work is partly supported by Fok Ying Tung Education Foundation under Grant No. 171006 and Natural Science Foundation of Jiangsu Province under Grant No.BK20211601.
\end{acknowledgments}

\bibliography{ShadowInCG20220408}

\begin{thebibliography}{99}
\baselineskip=0.5 cm


\bibitem{Bardeen1973}
Bardeen, J. M.,
``Les Houches Summer School of Theoretical Physics: Black Holes,''
Gordon and Breach Science Publishers, Inc., United States. (1973): 215-240

\bibitem{Synge:1966okc}
J.~L.~Synge,
``The Escape of Photons from Gravitationally Intense Stars,''
Mon. Not. Roy. Astron. Soc. \textbf{131} (1966) no.3, 463-466

\bibitem{Falcke:1999pj}
H.~Falcke, F.~Melia and E.~Agol,
``Viewing the shadow of the black hole at the galactic center,''
Astrophys. J. Lett. \textbf{528} (2000), L13
[arXiv:astro-ph/9912263 [astro-ph]].

\bibitem{Virbhadra:1999nm}
K.~S.~Virbhadra and G.~F.~R.~Ellis,
``Schwarzschild black hole lensing,''
Phys. Rev. D \textbf{62} (2000), 084003
[arXiv:astro-ph/9904193 [astro-ph]].

\bibitem{Shen:2005cw}
Z.~Q.~Shen, K.~Y.~Lo, M.~C.~Liang, P.~T.~P.~Ho and J.~H.~Zhao,
``A size of \textasciitilde{}1 au for the radio source sgr a* at the centre of the milky way,''
Nature \textbf{438} (2005), 62
[arXiv:astro-ph/0512515 [astro-ph]].

\bibitem{Younsi:2016azx}
Z.~Younsi, A.~Zhidenko, L.~Rezzolla, R.~Konoplya and Y.~Mizuno,
``New method for shadow calculations: Application to parametrized axisymmetric black holes,''
Phys. Rev. D \textbf{94} (2016) no.8, 084025
[arXiv:1607.05767 [gr-qc]].

\bibitem{Atamurotov:2013sca}
F.~Atamurotov, A.~Abdujabbarov and B.~Ahmedov,
``Shadow of rotating non-Kerr black hole,''
Phys. Rev. D \textbf{88} (2013) no.6, 064004

\bibitem{Atamurotov:2015xfa}
F.~Atamurotov, S.~G.~Ghosh and B.~Ahmedov,
``Horizon structure of rotating Einstein\textendash{}Born\textendash{}Infeld black holes and shadow,''
Eur. Phys. J. C \textbf{76} (2016) no.5, 273
[arXiv:1506.03690 [gr-qc]].






\bibitem{Amir:2017slq}
M.~Amir, B.~P.~Singh and S.~G.~Ghosh,
``Shadows of rotating five-dimensional charged EMCS black holes,''
Eur. Phys. J. C \textbf{78} (2018) no.5, 399
[arXiv:1707.09521 [gr-qc]].

\bibitem{Eiroa:2017uuq}
E.~F.~Eiroa and C.~M.~Sendra,
``Shadow cast by rotating braneworld black holes with a cosmological constant,''
Eur. Phys. J. C \textbf{78} (2018) no.2, 91
[arXiv:1711.08380 [gr-qc]].

\bibitem{Vagnozzi:2019apd}
S.~Vagnozzi and L.~Visinelli,
``Hunting for extra dimensions in the shadow of M87*,''
Phys. Rev. D \textbf{100} (2019) no.2, 024020
[arXiv:1905.12421 [gr-qc]].


\bibitem{Long:2019nox}
F.~Long, J.~Wang, S.~Chen and J.~Jing,
``Shadow of a rotating squashed Kaluza-Klein black hole,''
JHEP \textbf{10} (2019), 269
[arXiv:1906.04456 [gr-qc]].

\bibitem{Long:2020wqj}
F.~Long, S.~Chen, M.~Wang and J.~Jing,
``Shadow of a disformal Kerr black hole in quadratic degenerate higher-order scalar\textendash{}tensor theories,''
Eur. Phys. J. C \textbf{80} (2020) no.12, 1180
[arXiv:2009.07508 [gr-qc]].

\bibitem{Banerjee:2019nnj}
I.~Banerjee, S.~Chakraborty and S.~SenGupta,
``Silhouette of M87*: A New Window to Peek into the World of Hidden Dimensions,''
Phys. Rev. D \textbf{101} (2020) no.4, 041301
[arXiv:1909.09385 [gr-qc]].

\bibitem{Mishra:2019trb}
A.~K.~Mishra, S.~Chakraborty and S.~Sarkar,
``Understanding photon sphere and black hole shadow in dynamically evolving spacetimes,''
Phys. Rev. D \textbf{99} (2019) no.10, 104080
[arXiv:1903.06376 [gr-qc]].


\bibitem{Kumar:2020hgm}
R.~Kumar, S.~G.~Ghosh and A.~Wang,
``Gravitational deflection of light and shadow cast by rotating Kalb-Ramond black holes,''
Phys. Rev. D \textbf{101} (2020) no.10, 104001
[arXiv:2001.00460 [gr-qc]].

\bibitem{Qian:2021qow}
W.~L.~Qian, S.~Chen, C.~G.~Shao, B.~Wang and R.~H.~Yue,
``Cuspy and fractured black hole shadows in a toy model with axisymmetry,''
Eur. Phys. J. C \textbf{82} (2022) no.1, 91
[arXiv:2102.03820 [gr-qc]].

\bibitem{Zeng:2020dco}
X.~X.~Zeng, H.~Q.~Zhang and H.~Zhang,
``Shadows and photon spheres with spherical accretions in the four-dimensional Gauss\textendash{}Bonnet black hole,''
Eur. Phys. J. C \textbf{80} (2020) no.9, 872
[arXiv:2004.12074 [gr-qc]].

\bibitem{Zeng:2021dlj}
X.~X.~Zeng, G.~P.~Li and K.~J.~He,
``The shadows and observational appearance of a noncommutative black hole surrounded by various profiles of accretions,''
Nucl. Phys. B \textbf{974} (2022), 115639
[arXiv:2106.14478 [hep-th]].



\bibitem{Lin:2022ksb}
F.~L.~Lin, A.~Patel and H.~Y.~Pu,
``Black Hole Shadow with Soft Hair,''
[arXiv:2202.13559 [gr-qc]].


\bibitem{Sun:2022wya}
C.~Sun, Y.~Liu, W.~L.~Qian and R.~Yue,
``Shadows of magnetically charged rotating black holes surrounded by quintessence,''
[arXiv:2201.01890 [gr-qc]].


\bibitem{Cimdiker:2021cpz}
\.I.~\c{C}imdiker, D.~Demir and A.~\"Ovg\"un,
``Black hole shadow in symmergent gravity,''
Phys. Dark Univ. \textbf{34} (2021), 100900
[arXiv:2110.11904 [gr-qc]].


\bibitem{Zhong:2021mty}
Z.~Zhong, Z.~Hu, H.~Yan, M.~Guo and B.~Chen,
``QED effects on Kerr black hole shadows immersed in uniform magnetic fields,''
Phys. Rev. D \textbf{104} (2021) no.10, 104028
[arXiv:2108.06140 [gr-qc]].

\bibitem{Hou:2021okc}
Y.~Hou, M.~Guo and B.~Chen,
``Revisiting the shadow of braneworld black holes,''
Phys. Rev. D \textbf{104} (2021) no.2, 024001
[arXiv:2103.04369 [gr-qc]].


\bibitem{Cai:2021uov}
X.~C.~Cai and Y.~G.~Miao,
``Can we know about black hole thermodynamics through shadows?,''
[arXiv:2107.08352 [gr-qc]].


\bibitem{Gan:2021pwu}
Q.~Gan, P.~Wang, H.~Wu and H.~Yang,
``Photon spheres and spherical accretion image of a hairy black hole,''
Phys. Rev. D \textbf{104} (2021) no.2, 024003
[arXiv:2104.08703 [gr-qc]].


\bibitem{Chang:2021ngy}
Z.~Chang and Q.~H.~Zhu,
``The observer-dependent shadow of the Kerr black hole,''
JCAP \textbf{09} (2021), 003
[arXiv:2104.14221 [gr-qc]].


\bibitem{Wang:2021ara}
M.~Wang, S.~Chen and J.~Jing,
``Kerr black hole shadows in Melvin magnetic field with stable photon orbits,''
Phys. Rev. D \textbf{104} (2021) no.8, 084021
[arXiv:2104.12304 [gr-qc]].



\bibitem{Shaikh:2021cvl}
R.~Shaikh, S.~Paul, P.~Banerjee and T.~Sarkar,
``Shadows and thin accretion disk images of the $\gamma$-metric,''
[arXiv:2105.12057 [gr-qc]].




\bibitem{Guo:2020blq}
H.~Guo, H.~Liu, X.~M.~Kuang and B.~Wang,
``Acoustic black hole in Schwarzschild spacetime: quasi-normal modes, analogous Hawking radiation and shadows,''
Phys. Rev. D \textbf{102} (2020), 124019
[arXiv:2007.04197 [gr-qc]];
``Shadow and near-horizon characteristics of the acoustic charged black hole in curved spacetime,
''Phys. Rev. D \textbf{104} (2021) no.10, 104003
[arXiv:2107.05171 [gr-qc]]


\bibitem{Hioki:2009na}
K.~Hioki and K.~i.~Maeda,
``Measurement of the Kerr Spin Parameter by Observation of a Compact Object's Shadow,''
Phys. Rev. D \textbf{80} (2009), 024042
[arXiv:0904.3575 [astro-ph.HE]].

\bibitem{Kumar:2018ple}
R.~Kumar and S.~G.~Ghosh,
``Black Hole Parameter Estimation from Its Shadow,''
Astrophys. J. \textbf{892} (2020), 78
[arXiv:1811.01260 [gr-qc]].

\bibitem{Badia:2021kpk}
J.~Bada and E.~F.~Eiroa,
``Shadow of axisymmetric, stationary, and asymptotically flat black holes in the presence of plasma,''
Phys. Rev. D \textbf{104} (2021) no.8, 084055
[arXiv:2106.07601 [gr-qc]].

\bibitem{Wei:2013kza}
S.~W.~Wei and Y.~X.~Liu,
``Observing the shadow of Einstein-Maxwell-Dilaton-Axion black hole,''
JCAP \textbf{11} (2013), 063
[arXiv:1311.4251 [gr-qc]].

\bibitem{Allahyari:2019jqz}
A.~Allahyari, M.~Khodadi, S.~Vagnozzi and D.~F.~Mota,
``Magnetically charged black holes from non-linear electrodynamics and the Event Horizon Telescope,''
JCAP \textbf{02} (2020), 003
[arXiv:1912.08231 [gr-qc]].

\bibitem{Tsupko:2017rdo}
O.~Y.~Tsupko,
``Analytical calculation of black hole spin using deformation of the shadow,''
Phys. Rev. D \textbf{95} (2017) no.10, 104058
[arXiv:1702.04005 [gr-qc]].

\bibitem{Cunha:2019dwb}
P.~V.~P.~Cunha, C.~A.~R.~Herdeiro and E.~Radu,
``Spontaneously Scalarized Kerr Black Holes in Extended Scalar-Tensor\textendash{}Gauss-Bonnet Gravity,''
Phys. Rev. Lett. \textbf{123} (2019) no.1, 011101
[arXiv:1904.09997 [gr-qc]].





\bibitem{Kumar:2020owy}
R.~Kumar and S.~G.~Ghosh,
``Rotating black holes in $4D$ Einstein-Gauss-Bonnet gravity and its shadow,''
JCAP \textbf{07} (2020), 053
[arXiv:2003.08927 [gr-qc]].


\bibitem{Chen:2020aix}
C.~Y.~Chen,
``Rotating black holes without $\mathbb{Z}_2$ symmetry and their shadow images,''
JCAP \textbf{05} (2020), 040
[arXiv:2004.01440 [gr-qc]].

\bibitem{Brahma:2020eos}
S.~Brahma, C.~Y.~Chen and D.~h.~Yeom,
``Testing Loop Quantum Gravity from Observational Consequences of Nonsingular Rotating Black Holes,''
Phys. Rev. Lett. \textbf{126} (2021) no.18, 181301
[arXiv:2012.08785 [gr-qc]].

\bibitem{Belhaj:2020kwv}
A.~Belhaj, M.~Benali, A.~E.~Balali, W.~E.~Hadri and H.~El Moumni,
``Shadows of Charged and Rotating Black Holes with a Cosmological Constant,''
[arXiv:2007.09058 [gr-qc]].


\bibitem{Lee:2021sws}
B.~H.~Lee, W.~Lee and Y.~S.~Myung,
``Shadow cast by a rotating black hole with anisotropic matter,''
Phys. Rev. D \textbf{103} (2021) no.6, 064026
[arXiv:2101.04862 [gr-qc]].

\bibitem{Badia:2020pnh}
J.~Bad\'\i{}a and E.~F.~Eiroa,
``Influence of an anisotropic matter field on the shadow of a rotating black hole,''
Phys. Rev. D \textbf{102} (2020) no.2, 024066
[arXiv:2005.03690 [gr-qc]].

\bibitem{Frion:2021jse}
E.~Frion, L.~Giani and T.~Miranda,
``Black Hole Shadow Drift and Photon Ring Frequency Drift,''
[arXiv:2107.13536 [gr-qc]].

\bibitem{Roy:2021uye}
R.~Roy, S.~Vagnozzi and L.~Visinelli,
``Superradiance evolution of black hole shadows revisited,''
[arXiv:2112.06932 [astro-ph.HE]].

\bibitem{Afrin:2021imp}
M.~Afrin, R.~Kumar and S.~G.~Ghosh,
``Parameter estimation of hairy Kerr black holes from its shadow and constraints from M87*,''
Mon. Not. Roy. Astron. Soc. \textbf{504} (2021), 5927-5940
[arXiv:2103.11417 [gr-qc]].





\bibitem{Kumar:2019pjp}
R.~Kumar, S.~G.~Ghosh and A.~Wang,
``Shadow cast and deflection of light by charged rotating regular black holes,''
Phys. Rev. D \textbf{100} (2019) no.12, 124024
[arXiv:1912.05154 [gr-qc]].



\bibitem{Jha:2021bue}
S.~K.~Jha and A.~Rahaman,
``Study of shadow and parameter estimation of non-commutative Kerr-like Lorentz violating black holes,''
[arXiv:2111.02817 [gr-qc]].



\bibitem{Ghosh:2020spb}
S.~G.~Ghosh, R.~Kumar and S.~U.~Islam,
``Parameters estimation and strong gravitational lensing of nonsingular Kerr-Sen black holes,''
JCAP \textbf{03} (2021), 056
[arXiv:2011.08023 [gr-qc]].

\bibitem{Bambi:2019tjh}
C.~Bambi, K.~Freese, S.~Vagnozzi and L.~Visinelli,
``Testing the rotational nature of the supermassive object M87* from the circularity and size of its first image,''
Phys. Rev. D \textbf{100} (2019) no.4, 044057
[arXiv:1904.12983 [gr-qc]].



\bibitem{Khodadi:2021gbc}
M.~Khodadi, G.~Lambiase and D.~F.~Mota,
``No-hair theorem in the wake of Event Horizon Telescope,''
JCAP \textbf{09} (2021), 028
[arXiv:2107.00834 [gr-qc]].

\bibitem{Afrin:2021wlj}
M.~Afrin and S.~G.~Ghosh,
``Constraining rotating black holes in Horndeski theory with EHT observations of M87*,''
[arXiv:2110.05258 [gr-qc]].





\bibitem{Cunha:2018acu}
P.~V.~P.~Cunha and C.~A.~R.~Herdeiro,
``Shadows and strong gravitational lensing: a brief review,''
Gen. Rel. Grav. \textbf{50} (2018) no.4, 42
[arXiv:1801.00860 [gr-qc]].

\bibitem{Perlick:2021aok}
V.~Perlick and O.~Y.~Tsupko,
``Calculating black hole shadows: Review of analytical studies,''
Phys. Rept. \textbf{947} (2022), 1-39
[arXiv:2105.07101 [gr-qc]].




\bibitem{EventHorizonTelescope:2019dse}
K.~Akiyama \textit{et al.} [Event Horizon Telescope],
``First M87 Event Horizon Telescope Results. I. The Shadow of the Supermassive Black Hole,''
Astrophys. J. Lett. \textbf{875} (2019), L1
[arXiv:1906.11238 [astro-ph.GA]].

\bibitem{EventHorizonTelescope:2019ths}
K.~Akiyama \textit{et al.} [Event Horizon Telescope],
``First M87 Event Horizon Telescope Results. IV. Imaging the Central Supermassive Black Hole,''
Astrophys. J. Lett. \textbf{875} (2019) no.1, L4
[arXiv:1906.11241 [astro-ph.GA]].

\bibitem{EventHorizonTelescope:2019pgp}
K.~Akiyama \textit{et al.} [Event Horizon Telescope],
``First M87 Event Horizon Telescope Results. V. Physical Origin of the Asymmetric Ring,''
Astrophys. J. Lett. \textbf{875} (2019) no.1, L5
[arXiv:1906.11242 [astro-ph.GA]].

\bibitem{EventHorizonTelescope:2021dqv}
P.~Kocherlakota \textit{et al.} [Event Horizon Telescope],
``Constraints on black-hole charges with the 2017 EHT observations of M87*,''
Phys. Rev. D \textbf{103} (2021) no.10, 104047
doi:10.1103/PhysRevD.103.104047
[arXiv:2105.09343 [gr-qc]].

\bibitem{Cunha:2019ikd}
P.~V.~P.~Cunha, C.~A.~R.~Herdeiro and E.~Radu,
``EHT constraint on the ultralight scalar hair of the M87 supermassive black hole,''
Universe \textbf{5} (2019) no.12, 220
[arXiv:1909.08039 [gr-qc]].

\bibitem{Khodadi:2020jij}
M.~Khodadi, A.~Allahyari, S.~Vagnozzi and D.~F.~Mota,
``Black holes with scalar hair in light of the Event Horizon Telescope,''
JCAP \textbf{09} (2020), 026
[arXiv:2005.05992 [gr-qc]].

\bibitem{Liu:2012xn}
H.~S.~Liu and H.~Lu,
``Charged Rotating AdS Black Hole and Its Thermodynamics in Conformal Gravity,''
JHEP \textbf{02} (2013), 139
[arXiv:1212.6264 [hep-th]].

\bibitem{Weyl:1918pdp}
H.~Weyl,
``Reine Infinitesimalgeometrie,''
Math. Z. \textbf{2} (1918) no.3-4, 384-411

\bibitem{Mannheim:1988dj}
P.~D.~Mannheim and D.~Kazanas,
``Exact Vacuum Solution to Conformal Weyl Gravity and Galactic Rotation Curves,''
Astrophys. J. \textbf{342} (1989), 635-638

\bibitem{Varieschi:2009vlp}
G.~U.~Varieschi,
``A Kinematical Approach to Conformal Cosmology,''
Gen. Rel. Grav. \textbf{42} (2010), 929-974
[arXiv:0809.4729 [gr-qc]].

\bibitem{tHooft:2010xlr}
G.~'t Hooft,
``Probing the small distance structure of canonical quantum gravity using the conformal group,''
[arXiv:1009.0669 [gr-qc]].

\bibitem{tHooft:2014swy}
G.~'t Hooft,
``Local Conformal Symmetry: the Missing Symmetry Component for Space and Time,''
[arXiv:1410.6675 [gr-qc]].

\bibitem{Bender:2007wu}
C.~M.~Bender and P.~D.~Mannheim,
``No-ghost theorem for the fourth-order derivative Pais-Uhlenbeck oscillator model,''
Phys. Rev. Lett. \textbf{100} (2008), 110402
[arXiv:0706.0207 [hep-th]].

\bibitem{Mannheim:2000ka}
P.~D.~Mannheim and A.~Davidson,
``Fourth order theories without ghosts,''
[arXiv:hep-th/0001115 [hep-th]].

\bibitem{Mannheim:2011ds}
P.~D.~Mannheim,
``Making the Case for Conformal Gravity,''
Found. Phys. \textbf{42} (2012), 388-420
[arXiv:1101.2186 [hep-th]].

\bibitem{Maldacena:2011mk}
J.~Maldacena,
``Einstein Gravity from Conformal Gravity,''
[arXiv:1105.5632 [hep-th]].

\bibitem{Mureika:2016efo}
J.~R.~Mureika and G.~U.~Varieschi,
``Black hole shadows in fourth-order conformal Weyl gravity,''
Can. J. Phys. \textbf{95} (2017) no.12, 1299-1306
[arXiv:1611.00399 [gr-qc]].

\bibitem{Carter:1968rr}
B.~Carter,
``Global structure of the Kerr family of gravitational fields,''
Phys. Rev. \textbf{174} (1968), 1559-1571

\bibitem{Grenzebach:2014fha}
A.~Grenzebach, V.~Perlick and C.~L\"ammerzahl,
``Photon Regions and Shadows of Kerr-Newman-NUT Black Holes with a Cosmological Constant,''
Phys. Rev. D \textbf{89} (2014) no.12, 124004
[arXiv:1403.5234 [gr-qc]].

\bibitem{Tsukamoto:2017fxq}
N.~Tsukamoto,
``Black hole shadow in an asymptotically-flat, stationary, and axisymmetric spacetime: The Kerr-Newman and rotating regular black holes,''
Phys. Rev. D \textbf{97} (2018) no.6, 064021
[arXiv:1708.07427 [gr-qc]].

\bibitem{Xavier:2020egv}
S.~V.~M.~C.~B.~Xavier, P.~Cunha, V.P., L.~C.~B.~Crispino and C.~A.~R.~Herdeiro,
``Shadows of charged rotating black holes: Kerr\textendash{}Newman versus Kerr\textendash{}Sen,''
Int. J. Mod. Phys. D \textbf{29} (2020) no.11, 2041005
[arXiv:2003.14349 [gr-qc]].

\bibitem{Cunningham:cgh}
C.~T.~Cunningham, and J.~M.~Bardeen,
``The optical appearance of a star orbiting an extreme Kerr black hole,''
Astrophy. J. Lett. 173 (1972) L137

\bibitem{Cunha:2015yba}
P.~V.~P.~Cunha, C.~A.~R.~Herdeiro, E.~Radu and H.~F.~Runarsson,
``Shadows of Kerr black holes with scalar hair,''
Phys. Rev. Lett. \textbf{115} (2015) no.21, 211102
[arXiv:1509.00021 [gr-qc]].

\bibitem{Fathi:2020agx}
M.~Fathi, M.~Olivares and J.~R.~Villanueva,
``Ergosphere, Photon Region Structure, and the Shadow of a Rotating Charged Weyl Black Hole,''
Galaxies \textbf{9} (2021) no.2, 43
[arXiv:2011.04508 [gr-qc]].

\bibitem{Ghosh:2020ece}
S.~G.~Ghosh, M.~Amir and S.~D.~Maharaj,
``Ergosphere and shadow of a rotating regular black hole,''
Nucl. Phys. B \textbf{957} (2020), 115088
[arXiv:2006.07570 [gr-qc]].

\bibitem{CraigWalker:2018vam}
R.~Craig Walker, P.~E.~Hardee, F.~B.~Davies, C.~Ly and W.~Junor,
``The Structure and Dynamics of the Subparsec Jet in M87 Based on 50 VLBA Observations over 17 Years at 43 GHz,''
Astrophys. J. \textbf{855} (2018) no.2, 128
[arXiv:1802.06166 [astro-ph.HE]].

\bibitem{Perlick:2004tq}
V.~Perlick,
``Gravitational lensing from a spacetime perspective,''
Living Rev. Rel. \textbf{7} (2004), 9

\bibitem{Hasse:2005vu}
W.~Hasse and V.~Perlick,
``A Morse-theoretical analysis of gravitational lensing by a Kerr-Newman black hole,''
J. Math. Phys. \textbf{47} (2006), 042503
[arXiv:gr-qc/0511135 [gr-qc]].


\end{thebibliography}

\end{document}